\documentclass{jfm}
\usepackage{amsmath}
\usepackage{stackengine}
\usepackage{pgfplots}
\usepackage{tcolorbox} 
\usepackage{todonotes}
\usepackage{graphicx}
\usepackage{braket}
\usepackage{physics}
\usepackage[percent]{overpic}
\usepackage{commath}
\usepackage{caption}
\usepackage[crop=pdfcrop]{pstool}
\usepackage[position=top]{subcaption}
\DeclareMathOperator{\sign}{sign}
\usepackage{booktabs} 
\usepackage{bm}

\usepackage{tikz}
\usepackage[utf8]{inputenc}
\usepackage{pgfplots} 
\usepackage{pgfgantt}
\usepackage{pdflscape}
\pgfplotsset{compat=newest} 
\pgfplotsset{plot coordinates/math parser=false}

\newlength\figureheight
\newlength\figurewidth
\pgfplotsset{compat=newest}
\pgfplotsset{try min ticks=6}

\title{Hydrodynamic forces on assemblies of non-spherical particles: orientation and voidage effects}

\shorttitle{Hydrodynamic forces on assemblies of non-spherical particles}
\shortauthor{S. K. P. Sanjeevi and J. T. Padding}

\author{Sathish K. P. Sanjeevi
 \and Johan T. Padding
 \corresp{\email{J.T.Padding@tudelft.nl}}
 }

\affiliation{
Process and Energy Department, Delft University of Technology, Leeghwaterstraat 39, 
\newline 2628 CB Delft, The Netherlands
}

\begin{document}

\maketitle

\begin{abstract}
This work provides a recipe for creating drag, lift and torque closures for static assemblies of axisymmetric, non-spherical particles. Apart from Reynolds number $\Rey$ and solids volume fraction $\epsilon_s$, we propose four additional parameters to characterize the flow through non-spherical particle assemblies. Two parameters consider the mutual orientations of particles (the orientation tensor eigenvalues $S_1$ and $S_2$) and two angles represent the flow direction (polar and azimuthal angles $\alpha$ and $\beta$). Interestingly, we observe that the hydrodynamic forces on the particles are independent of the mutual particle orientations. Rather, the most important parameter representing the particle configuration itself is the incident angle $\phi$ of the individual particles with respect to the incoming flow. Moreover, we observe that our earlier finding of sine-squared scaling of drag for isolated particles \citep{sanjeevi2017orientational} holds on average even for a multiparticle system in both the viscous and inertial regimes. Similarly, we observe that the average lift for a multiparticle system follows sine-cosine scaling, as is observed for isolated particles. Such findings are very helpful since the pressure drop of a packed bed or porous media can be computed just with the knowledge of orientation distribution of particles and their drag at $\phi=0^\circ$ and $\phi=90^\circ$ for a given $\Rey$ and $\epsilon_s$. With the identified dependent parameters, we propose drag, lift and torque closures for multiparticle systems.
\end{abstract}


\section{Introduction}
Accurate fluid-particle drag, lift and torque closures are required for precise Euler-Lagrangian simulations of non-spherical particles. Historically, different drag closures have been developed for assemblies of spherical particles \citep{beetstra2007drag,tenneti2011drag, yali2015new}. However, practical flows often involve assemblies of non-spherical particles for which there exist no closures at the moment. Even for static, mono-disperse, non-spherical particle assemblies, creating the required closures is complicated due to the different possible mutual orientations of the particles. Furthermore, there is a lack of knowledge identifying the relevant parameters that can parametrize the drag, lift and torque, which adds to the complication. Most fluidization applications involve gas-solid flows, in which case the large density ratios ensure large Stokes numbers, i.e. the typical relaxation time of the solid particle velocity is large relative to the response time of the gas \citep{sanjeevi2018drag}. It has been shown that under such conditions, it is sufficient to assume the particle configurations to be quasi-static \citep{rubinstein2017lattice}.

Conventionally, fluidization simulations of non-spherical particles are performed by combining isolated particle drag correlations with correlations expressing the voidage effects as determined for sphere assemblies. There have been several works in the past focussing on the drag experienced by isolated non-spherical particles. \citet{holzer2008new} proposed a correlation for the drag coefficient $C_D$ for non-spherical particles. The proposed correlation is a function of particle sphericity and crosswise-sphericity, based on the projected area, which indirectly represents the particle orientation. Their proposed correlation is based on literature data of different non-spherical particles of various shapes and aspect ratios. More recently, drag, lift and torque closures for isolated non-spherical particles have been derived based on direct numerical simulations. \citet{zastawny2012derivation} developed drag, lift and torque coefficients for four different non-spherical particles as a function of Reynolds number $\Rey$ and incident angle $\phi$ with respect to the incoming flow. The investigated particles have aspect ratios ranging from 1.25 to 5 and $\Rey\leq300$. Similarly, \citet{richter2013new} proposed fits for drag, lift, torque coefficients for cubic and ellipsoidal particles. The above mentioned literature is primarily limited to steady flow conditions. Recently, we developed drag, lift and torque closures for three different non-spherical particles from the viscous Stokes regime upto the high $\Rey$ regime of $\Rey=2000$, involving complex, unsteady flows \citep{sanjeevi2018drag}. In an earlier work \citep{sanjeevi2017orientational}, we reported the interesting finding that the drag coefficient $C_{D}$ at different incident angles $\phi$ follows a sine-squared scaling given by
\begin{eqnarray}
C_{D,\phi} = C_{D,\phi=0^{\circ}}+(C_{D,\phi=90^{\circ}}-C_{D,\phi=0^{\circ}})\sin^2\phi.
\label{eq:cdStokes}
\end{eqnarray}
Likewise, we reported another interesting finding that the lift coefficient $C_L$ follows sine-cosine scaling at different $\phi$ as
\begin{eqnarray}
C_{L,\phi}=(C_{D,\phi=90^{\circ}}-C_{D,\phi=0^{\circ}})\sin\phi \cos\phi
\label{eq:CL_stokes}
\end{eqnarray}
for various elongated particles. The above mentioned scaling laws must be mathematically true in the Stokes regime due to linearity of the flow fields. However, their validity in the inertial regimes is primarily due to an interesting pattern of pressure distribution contributing to the drag and lift for different incident angles \citep{sanjeevi2017orientational}. In equations \ref{eq:cdStokes} and \ref{eq:CL_stokes}, the drag coefficients at incident angles of 0 and 90 degrees still depend on particle shape and Reynolds number. The Reynolds number in the present work is defined as $\Rey=|\bm{u}_s|d_{eq}/\nu$, where $\bm{u}_s$ is the superficial flow velocity, $\nu$ is the kinematic viscosity of the fluid, and $d_{eq}$ is the diameter of the volume-equivalent sphere given by $d_{eq}=(6V_p/\pi)^{1/3}$ with $V_p$ the particle volume.

For multiparticle systems, various literature is available to include the voidage effects, often developed through experiments and numerical simulations. One of the most widely used expressions is that of \citet{ergun1952fluid}, which has been developed based on a series of packed bed experiments of different particle shapes. The only limitation of this work is that it is applicable primarily in the dense limit. \citet{richardson1954sedimentation} performed various sedimentation and fluidization experiments and proposed accordingly the effect of particle volume fraction on the drag. Based on the previous literature on sedimentation and packed bed experiments, \citet{di1994voidage} bridged the dilute and dense particulate regimes through a unified function, which also extends from low to high $\Rey$. Though the above correlations provide a good approximation, the use of such closures in Euler-Lagrangian simulations often do not represent accurate physics. This is mainly due to the inability to construct moderate solids volume fractions in experiments.

There is a growing interest to use numerical simulations to accurately develop drag closures for different Reynolds numbers $\Rey$ and solids volume fractions $\epsilon_s$, albeit primarily for spheres. Initially, lattice Boltzmann method (LBM) has been the choice for simulating assemblies of spheres \citep{hill2001moderate, van2005lattice, beetstra2007drag}. Recently, \citet{tenneti2011drag} used an immersed boundary method (IBM) to develop drag closures for static assemblies of spheres for $0.01\leq\Rey\leq300$ and $0.1\leq\epsilon_s\leq0.5$. They observed a deviation of 30\% in the $\Rey$ range from 100 to 300 with respect to the earlier work of \citet{beetstra2007drag}. This is possible because \citet{beetstra2007drag} used LBM with the conventional stair-case boundary condition to represent the sphere boundaries, for which at high $\Rey$ thinner boundary layers result in larger deviations. In this work, we use a multi-relaxation time (MRT) LBM for high $\Rey$ flows and an interpolated bounceback scheme to much more accurately represent the particle geometry. Recently, \citet{yali2015new} used an IBM based solver to create drag closures for static assemblies of spheres upto $\Rey\leq1000$ and $\epsilon_s\leq0.6$.

There are several disadvantages with combining an isolated non-spherical particle drag with a voidage function based on spheres. First, the assumption that the voidage effects are independent of particle shape is probably incorrect, since there exist different closures even for assemblies of polydisperse spheres \citep{beetstra2007drag, holloway2010fluid}. Second, the voidage effects on lift and torque in a multiparticle system are unknown and hence are often neglected in Euler-Lagrangian simulations \citep{oschmann2014numerical, mahajan2018nonspherical}. Thirdly, using the same factor for voidage effects for all incident angles $\phi$ may hold in sufficiently dilute regimes but its validity in the dense limit is unknown. At the moment, only \citet{he2018variation} have discussed the drag, lift and torque for an assembly of non-spherical particles. However, they do not propose any correlations which can be used in Euler-Lagrangian simulations. This could be due to the difficulty in identifying the dependent parameters which represent the orientation effects in non-spherical, multiparticle system adequately.

In this work, we propose and subsequently identify the important dependent parameters for static, mono-disperse assemblies of axisymmetric non-spherical particles. With the identified parameters, we create the drag, lift and torque closures accordingly. Our particle of interest is a capsule-like spherocylinder of aspect ratio 4 (total length/shaft diameter). Compared to the two parameters for sphere assemblies, i.e. Reynolds number $\Rey$ and solids volume fraction $\epsilon_s$, we propose four additional parameters for the assembly of axisymmetric non-spherical particles. Two parameters describe the mutual orientations of the particles, namely two eigenvalues $S_1$ and $S_2$ of the orientation tensor, and two angle parameters $\alpha$ and $\beta$ represent the polar and azimuthal angles of the average flow (in the coordinate frame determined by the principal directions of the order tensor). The resulting six dimensional parameter space is adequately explored and correlations are proposed accordingly. It should be noted that the fixed nature of the particles in our simulations imply that the proposed correlations are applicable for high Stokes number flows as typically experienced by Geldart D category particles.

\section{Numerical method}
\subsection{Lattice Boltzmann method}
In the present work, we use a D3Q19, multi-relaxation time (MRT) lattice Boltzmann method \citep{d2002multiple} to simulate the fluid flow. The numerical method is adequately explained and validated in our previous works \citep{sanjeevi2017orientational, sanjeevi2018drag}. The evolution of particle distribution function $\ket{f}$ is computed as
\begin{equation}
\ket{f(\bm{r}+\bm{e}_{\alpha} \Delta t, t+\Delta t)} = \ket{f(\bm{r},t)} -\mathsfbi{M}^{-1}\mathsfbi{\hat{S}}(\ket{m(\bm{r},t)}-\ket*{m^{(eq)}(\bm{r},t)}) ,
\label{eq:lbe_equation}
\end{equation}
for position $\bm{r}$ with discrete velocities $\bm{e}_\alpha$ in directions $\alpha=1,2...,19$. Equation \ref{eq:lbe_equation} is solved in a sequence of two steps namely collision and streaming. $\mathsfbi{M}$ is a $19\times 19$ transformation matrix used to transform $\ket{f}$ from velocity space to moment space $\ket{m}$ with $\ket{m}=\mathsfbi{M}\cdot\ket{f}$. Here, the ket vector $\ket{\cdot}$ implies a column vector. The relaxation matrix $\mathsfbi{\hat{S}}=\mathsfbi{M}\cdot\mathsfbi{S}\cdot\mathsfbi{M}^{-1}$ is a $19\times 19$ diagonal matrix. $\mathsfbi{\hat{S}}$ utilizes different, optimally chosen relaxation rates for different moments, thereby providing better stability compared to the single-relaxation-time LBM scheme \citep{d2002multiple}. The matrices $\mathsfbi{M}$ and $\mathsfbi{\hat{S}}$ are similar to \citet{huang2012rotation} and are given in \citet{sanjeevi2018drag}. The density is computed as $\rho=\sum_{\alpha} f_\alpha$ and the momentum as $\rho \bm{u} = \sum_{\alpha}f_\alpha e_\alpha$. The relation between the kinematic viscosity of the fluid and the dimensionless relaxation time $\tau$ is $\nu=c_s^2(\tau-1/2)\Delta t$, and the pressure $p$ is related to the density by $p=\rho c_s^2$, where $c_s$ is the speed of sound. A linearly interpolated bounce back scheme \citep{bouzidi2001momentum, lallemand2003lattice} is used to accurately consider the curved geometry of the particle, as opposed to the traditional stair-case bounce back boundary condition. The flow is driven by a body force $\bm{g}$ and the simulated domain is periodic in all three directions. The use of the interpolated bounce back scheme within a periodic domain results in a slow mass leakage/gain in the system. Accordingly, the mass is corrected using a case 3 type correction described in \citet{sanjeevi2018choice}. The results for the multiparticle system are validated in section \ref{sec:multiparticle_validation}.

\begin{table}
\begin{center}
\def~{\hphantom{0}}
\begin{tabular}{cccc}
\toprule
$\Rey$                &$L_D$       & $d_{eq}$    &$\nu$\\
\midrule
$0.1\leq\Rey\leq10$   &288         &$28.36-48.5$   &1.3/3\\
$10<\Rey\leq100$      &576         &$56.72-97.0$   &0.1 to 0.08/3\\
300                   &576         &$56.72-97.0$   &0.04/3\\
600                   &576         &$56.72-97.0$   &0.015/3\\
1000                  &768         &$75.63-129.3$  &0.01/3\\
\bottomrule
\end{tabular}
\caption{Details of the simulation parameters used in our simulations in LB units. $L_D$ denotes the side length of the cubic domain. The range of $d_{eq}$ specified is respectively for $0.1\leq\epsilon_s\leq0.5$.}
\label{tab:simulation_parameters}
\end{center}
\end{table}
The ratio of $d_{eq}/d_{min}$ equals 1.765 for the considered spherocylinder of aspect ratio 4, where $d_{min}$ implies diameter of the cylinder. The simulation parameters used in our LBM simulations are summarized in table \ref{tab:simulation_parameters}. Specifically, it can observed that a good particle resolution ($d_{eq}$) is maintained for different $\Rey$. Further with increasing $\epsilon_s$, the $d_{eq}$ is increased accordingly to resolve increased velocity gradients at high $\epsilon_s$. All LBM simulations have cubic domain, each with 200 particles unless otherwise specified. At least two independent simulations are performed for each $\Rey$ and $\epsilon_s$ and the details of independent number of simulations are discussed later (see figure \ref{fig:multiRegimeMap}).

\subsection{Flow control}
In order to perform a simulation for a specific $\Rey$, it is required to control the superficial flow velocity $\bm{u}_s$ by applying a body force $\bm{g}$. The relationship between the superficial velocity and the average interstitial flow velocity $\bm{u}_{avg}$ is given by $\bm{u}_s=(1-\epsilon_s)\bm{u}_{avg}$. Due to the non-spherical nature of the particles, the sum of lift forces is often non-zero, and the resultant direction of $\bm{u}_s$ can be different from the direction of $\bm{g}$. This necessitates the need to control both direction and magnitude of the body force. Initially, the fluid is at rest with both $\bm{u}_s$ and $\bm{g}$ zero. The flow is slowly ramped up by increasing $\bm{g}$ until the desired $\bm{u}_s$ is achieved. For each timestep, the updated gravity $\bm{g}_{new}$ is computed as
\begin{eqnarray}
\bm{g}_{new} = \bm{g}_{prev}+\frac{(\bm{u}_{s,ref}-\bm{u}_{s,prev})}{K_p^2}\Delta t,
\end{eqnarray}
where $\bm{g}_{prev}$ is the gravity from the previous timestep, $\bm{u}_{s,ref}$ is the desired reference superficial velocity, and $\bm{u}_{s,prev}$ is the superficial velocity from the previous timestep. $K_p$ is a time constant which controls the system response rate. The stopping criterion for the simulations is when the system $\bm{u}_s$ reaches 99.9\% of the reference setpoint.

\section{Simulation setup}

\subsection{Orientation tensors}
In this section, we briefly explain the characterization of mutual orientations in an assembly of axisymmetric non-spherical particles with orientation tensors. We subsequently explain the use of a Maier-Saupe potential to achieve the desired particle configurations through Monte-Carlo simulations.
\begin{figure}
\centering
\psfragfig[width=\textwidth]{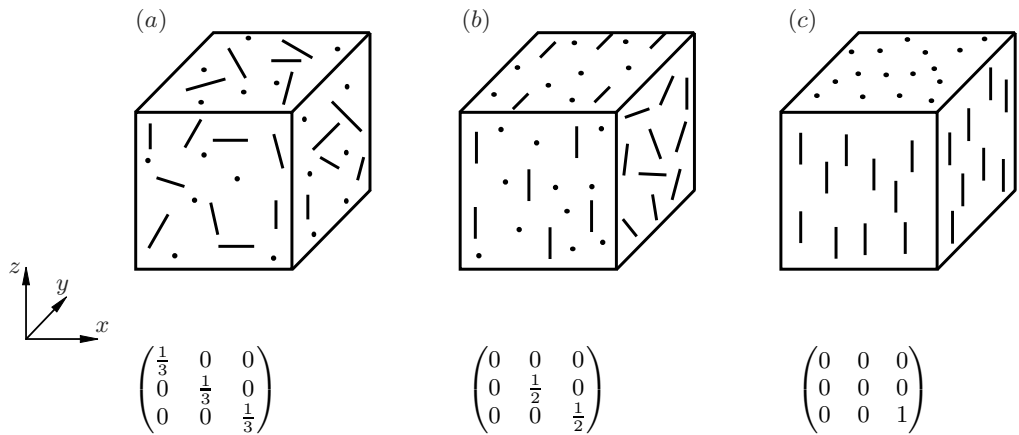}\\
\hspace*{35pt}
$\begin{pmatrix}
\frac{1}{3} &0 &0\\
0 &\frac{1}{3} &0\\
0 &0 &\frac{1}{3}
\end{pmatrix}$
\hspace*{65pt}
$\begin{pmatrix}
0 &0 &0\\
0 &\frac{1}{2} &0\\
0 &0 &\frac{1}{2}
\end{pmatrix}$
\hspace*{65pt}
$\begin{pmatrix}
0 &0 &0\\
0 &0 &0\\
0 &0 &1
\end{pmatrix}$\hspace*{20pt}
\caption{Different particle configurations and their orientation tensors: ($a$) Random, ($b$) planar random, and ($c$) unidirectional (nematic) configuration.}
\label{fig:Orientation_tensors}
\end{figure}

To describe the orientation of a single axisymmetric particle, the azimuthal and polar angles are sufficient. For a multiparticle configuration, it is important to parametrize the mutual orientations of the particles, with the least number of parameters. For this, we propose to use the orientation tensor $\mathsfbi{S}$ which is defined as the average of the dyadic products of the particle orientation vectors. In other words,
\begin{eqnarray}
\mathsfbi{S} = \left\langle\bm{p}\bm{p}^T\right\rangle. 
\end{eqnarray}
Here, $\bm{p}$ is the unit orientation vector of a particle. The 3 eigenvalues (which we order as $S_1, S_2, S_3$ from small to large) characterize the type of mutual alignment, as shown in figure \ref{fig:Orientation_tensors}. The  corresponding 3 eigenvectors define the principal directions of mutual particle alignment.

Because the trace of S is 1, only 2 eigenvalues are sufficient to specify the amount of randomness, planar random (bi-axial), or unidirectional (nematic) order. It should be noted that the tensor $\mathsfbi{S}$ is insensitive to an orientation $\bm{p}$ or $-\bm{p}$ of particles. In other words, the tensor captures essentially the mutual alignment of particles irrespective of particles oriented in positive or negative direction. Figure \ref{fig:Orientation_tensors}($a$) shows a completely random configuration with $S_1=S_2=S_3=1/3$. Figure \ref{fig:Orientation_tensors}($b$) shows a planar random configuration with particles primarily confined to planes (in this example with random orientations in planes normal to the $x$-direction) resulting in $S_1 = 0, S_2 = S_3 = 1/2$, and similarly a unidirectional (nematic, in this example in the $z$-direction) configuration in figure \ref{fig:Orientation_tensors}($c$) with $S_1 = S_2 =0, S_3 = 1$. In practical conditions, particles can exhibit more complex configurations in between these extremes but can be adequately described by 2 eigenvalues $S_1$ and $S_2$. Regarding the unidirectional case, we consider only nematic configurations but not smectic because ordering of both positions and orientations is rare in fluidization conditions.

The above metrics can be used to describe the particle configuration. However, due to the nonsphericity of the particles, the flow orientation with respect to the principal directions of the particle orientations is also important. This results in two parameters, namely the polar angle ($\alpha$) and azimuthal angle ($\beta$) of the average flow velocity vector with respect to the space spanned by the 3 eigenvectors of the orientation tensor. In summary, the parameter space to be explored for our flow problem has 6 parameters, namely Reynolds number $\Rey$, solids volume fraction $\epsilon_s$, two particle configuration parameters $S_1$, $S_2$ and two angles $\alpha$ and $\beta$ describing the mean flow orientation with respect to the configuration.

\subsection{Generation of biased particle configurations}
\label{sec:configurations}
\begin{figure}
\centering
\psfragfig[width=\textwidth]{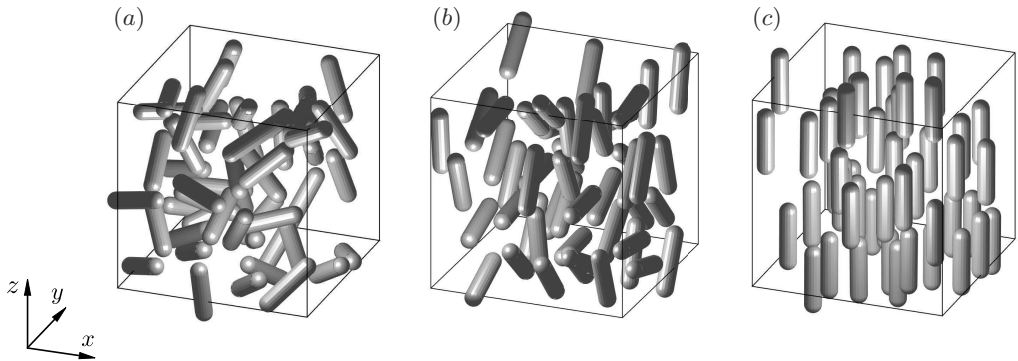}
\caption{Different configurations of non-spherical particles generated using the Monte-Carlo simulations: ($a$) Random configuration without the use of Maier-Saupe potential, ($b$) planar random, and ($c$) unidirectional configuration generated using the Maier-Saupe potential. For better clarity, the shown examples have only 50 particles. The actual simulations involve 200 particles.}
\label{fig:nonsph_configs}
\end{figure}
The generation of non-overlapping configurations of the particles in a periodic domain is required as an input for the flow simulations. Further, it is also required to generate configurations of particles with a prescribed orientation tensor, which adds further complexity. In this section, we briefly describe the Monte-Carlo simulation algorithm for generating configuration of non-overlapping particles and the use of a Maier-Saupe potential \citep{maier1959einfache} to bias the system to produce the required orientation tensor.

As the particles are spherocylindrical in shape, a simple way to detect overlap is to find the minimum distance between two line segments. We define the line segment as the line connecting the centres of the two spheres at the extremes of the spherocylinder. If the distance between two line segments is less than the particle diameter, then the spherocylinders overlap. A fast algorithm is used to measure the shortest distance between the line segments \citep{vega1994fast}. 

\begin{figure}
\centering
\psfragfig[width=\linewidth]{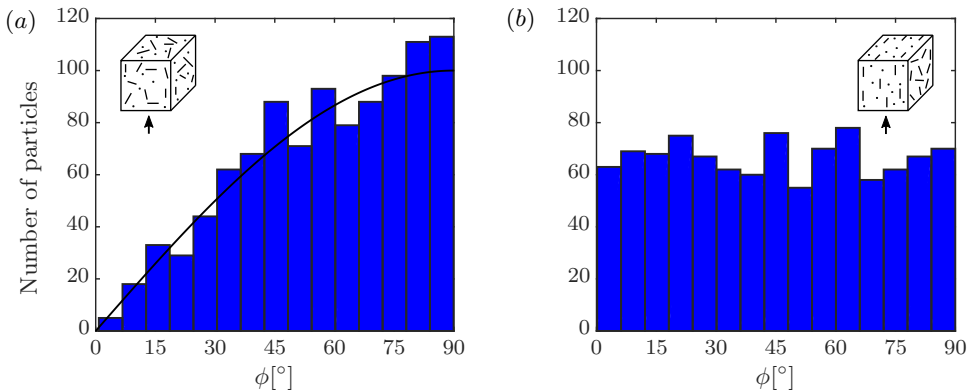}
\caption{Histogram of particles with different incident angles $\phi$ with respect to the flow vector (indicated by an arrow) for ($a$) random and ($b$) planar random configuration. The shown example has 1000 particles. It should be noted that the $\phi$ distribution for a random configuration will always scale as $\sin\phi$ (solid black curve) irrespective of the flow direction.}
\label{fig:particle_hist}
\end{figure}

Using the above overlap detection algorithm, the particles are randomly translated in small steps compared to the particle size and rotated by a small angle around a randomly chosen axis. This procedure results in a random configuration after many iterations. If a prescribed orientation tensor is required, besides the requirement of no overlap, the following Maier-Saupe potential is applied to accept or reject a new orientation of a particle. We define a director $\bm{n}$ along which the system is biased towards or against. Inside each Monte-Carlo simulation step, a new particle orientation $\bm{p}_{new}$ can be accepted or rejected from the current orientation $\bm{p}_{curr}$ based on following criteria:
\begin{eqnarray}
&\bm{p}_{new} =
\begin{cases}
    \bm{p}_{new},& \text{if } dE<0\\
    \bm{p}_{new},& \text{if } dE\geq0 \text{ and } U([0,1])< \exp(-dE)\\    
    \bm{p}_{curr},              & \text{otherwise} 
\end{cases}\\
\label{eq:MaierSaupeCondition}
\text{ where } &dE = A((\bm{p}_{new}\cdot\bm{n})^2 - (\bm{p}_{curr}\cdot\bm{n})^2).
\label{eq:diffPotential}
\end{eqnarray}
Here, $dE$ is the increase in Maier-Saupe potential and  $U([0,1])$ is a random number uniformly distributed between 0 and 1. The mutual particle orientations emerge from the balance between the random rotations, which tend to disorder the particle orientations, and the Maier-Saupe potential, which tend to order the particle orientations. The magnitude of $A$ determines the intensity of the configuration towards the director. A planar random configuration is achieved with the plane perpendicular to the director $\bm{n}$, if $A$ is positive. A unidirectional configuration along the direction of $\bm{n}$ is achieved, if $A$ is negative. Higher absolute $A$ values result in better perfection towards the desired configuration. With the mentioned strategy, any configuration in-between the ideal cases shown in figure \ref{fig:Orientation_tensors} can be achieved. Some sample configurations generated using the above mentioned algorithm are shown in figure \ref{fig:nonsph_configs}. For simplicity, the eigenvectors of the orientation tensor $\mathsfbi{S}$ are considered as aligned with the Cartesian coordinate system in figure \ref{fig:nonsph_configs}. The shown configurations are respectively equivalent to figure \ref{fig:Orientation_tensors}. For better clarity, the shown configuration has only 50 particles and the solids volume fraction $\epsilon_s$ is 0.1. The actual flow simulations have 200 particles and are performed for various $\epsilon_s$.


A common intuition may be that a random configuration would result in particles with evenly distributed values of the incident angle  $\phi$. However for a random configuration, the available number of particles at different $\phi$ are not uniform, as shown in figure \ref{fig:particle_hist}($a$). This is due to the higher probability to find particles at an angle $\phi$ near $90^\circ$ because the Jacobian for a spherical coordinate system scales as $\sin\phi$. Therefore, the disadvantage for a random configuration is that there are actually few data points at $\phi=0^\circ$ to create angle-dependent closures. On the contrary, the planar configuration with the planes parallel to the flow direction results in even particle distributions, as shown in figure \ref{fig:particle_hist}($b$). This information is considered while we generate configurations for the flow simulations.

\subsection{Forces and torques acting on a particle}
\begin{figure}
\centering
\psfragfig[width=0.4\linewidth]{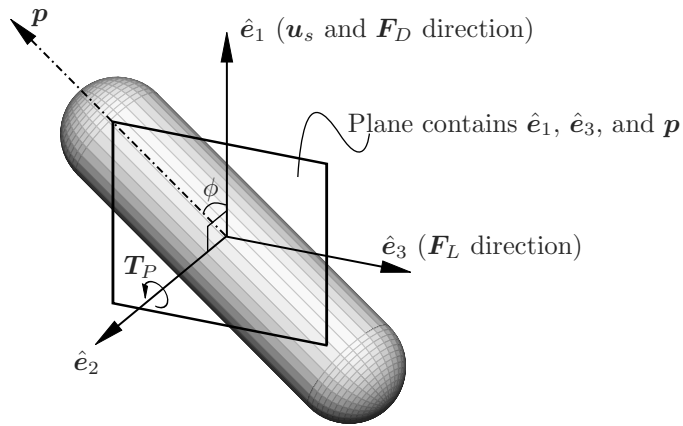}
\caption{The local coordinate system of a particle. $\bm{u}_s$ and $\bm{F}_D$ act along $\hat{\bm{e}}_1$, $\bm{F}_L$ along $\hat{\bm{e}}_3$ and $\bm{T}_P$ about the $\hat{\bm{e}}_2$ axis.}
\label{fig:lcs_particle}
\end{figure}

For an assembly of particles, different definitions are used to report the forces \citep{beetstra2007drag,tenneti2011drag,yali2015new}. To ensure consistency, it is important to know the form of the reported results. For a packed bed of particles in a flow induced by a macroscopic pressure gradient $\nabla P$, each particle of volume $V_p$ experiences a resulting force $\bm{F}$ due to the flow and a buoyancy force $\bm{F}_b=-V_p\nabla P$ due to the pressure gradient. For such a case, the total fluid-to-particle force $\bm{F}_{f\rightarrow p}$ acting on a particle is
\begin{eqnarray}
\bm{F}_{f\rightarrow p} = \bm{F}+\bm{F}_b.
\end{eqnarray}
Given $N$ particles with each of volume $V_p$ and total volume of the system $V$, the solids volume fraction is given by $\epsilon_s=NV_p/V$. 
Further, the relationship between $\bm{F}$ and $\bm{F}_{f\rightarrow p}$ is given by \citep{yali2015new}
\begin{eqnarray}
\bm{F} = \bm{F}_{f\rightarrow p}(1-\epsilon_s).
\end{eqnarray}

\begin{figure}
\centering
\psfragfig[width=0.5\textwidth]{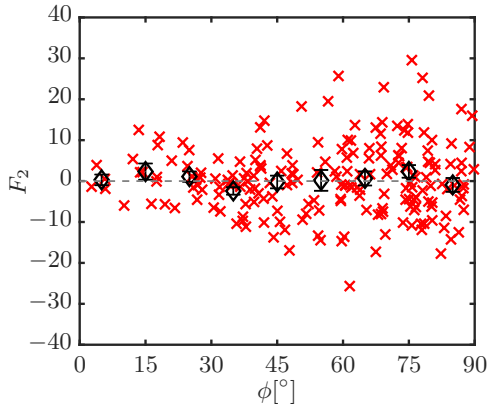}
\caption{Lateral force $F_2$ distribution for different particles ($\cross$) with averages at regular $\phi$ intervals ($\Diamond$) in a random configuration at $\Rey=100$ and $\epsilon_s=0.3$.}
\label{fig:force_independence}
\end{figure}

\begin{figure}
\centering
\psfragfig[width=\textwidth]{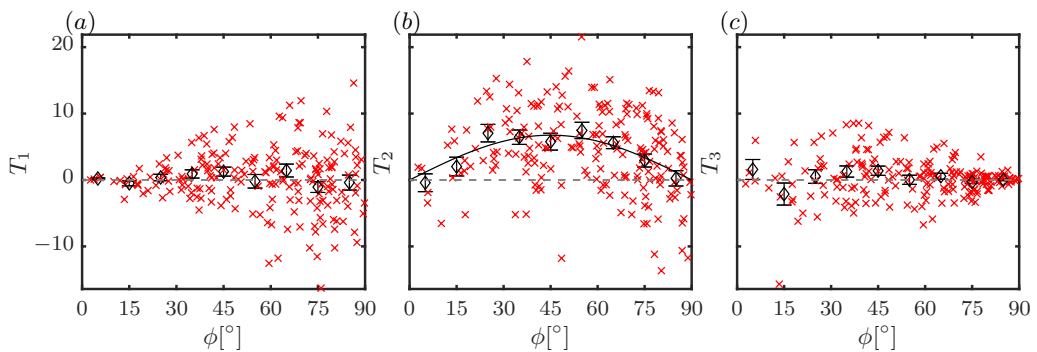}
\caption{Torques ($a$) $T_1$, ($b$) $T_2$, and ($c$) $T_3$ distribution for different particles ($\cross$) with averages at regular $\phi$ intervals ($\Diamond$) in a random configuration at $\Rey=100$ and $\epsilon_s=0.3$. Due to flow symmetry, the average $T_1$ and $T_3$ acting on particles are statistically zero. However, the pitching torque $T_P$ (or $T_2$) scales proportional to $\sin\phi\cos\phi$ (solid black line).}
\label{fig:torque_Indpendence}
\end{figure}

In this work, we report the forces $\bm{F}$ due to the flow and not $\bm{F}_{f\rightarrow p}$. The effects of buoyancy on torques are unknown and hence the reported torques $\bm{T}$ are also as they are determined from the simulations. We normalize the force and torque with the Stokes drag and torque of a volume-equivalent sphere:
\begin{eqnarray}
&\bm{F}_{norm}&=\frac{\bm{F}}{6\pi\mu R_{eq}|\bm{u}_s|}, \text{ and}\\
&\bm{T}_{norm}&=\frac{\bm{T}}{8\pi\mu R_{eq}^2|\bm{u}_s|}.
\end{eqnarray}
Here, $\mu$ is the dynamic viscosity and $R_{eq}$ is the radius of the volume equivalent sphere. Let $\bm{p}$ be the normalized orientation vector of the considered particle. The local coordinate system for each particle is defined as
\begin{eqnarray}
&\hat{\bm{e}}_1 &= \frac{\bm{u}_s}{|\bm{u}_s|},\\
&\hat{\bm{e}}_2 &= \frac{\hat{\bm{e}}_1 \cross \bm{p}}{|\hat{\bm{e}}_1 \cross \bm{p}|}\sign (\hat{\bm{e}}_1 \cdot \bm{p}), \text{ and}\\
&\hat{\bm{e}}_3 &= \hat{\bm{e}}_1 \cross \hat{\bm{e}}_2.
\end{eqnarray}
The above defined axes are accordingly illustrated in figure \ref{fig:lcs_particle}. The incident angle $\phi$ a particle makes with respect to the incoming flow is given by $\phi=\cos^{-1}(|\hat{\bm{e}}_1\cdot\bm{p}|)$. We also compute the average forces and torques for different $\phi$ intervals. Due to the finite number of measurements in these intervals, there is an error on the mean $\bar{x}$ of any property $x$. We use the standard error on the mean $\sigma_{\bar{x}}$ for the errorbars, computed as
\begin{eqnarray}
\sigma_{\bar{x}}=\sigma/\sqrt{n}.
\end{eqnarray}
Here $\sigma$ is the standard deviation of the corresponding variable $x$ and $n$ is the number of data points within the given $\phi$ interval. The normalized drag $F_D$ and lift $F_L$ can be computed from $\bm{F}_{norm}$ as
\begin{eqnarray}
F_D = &F_1 &= \bm{F}_{norm}\cdot \hat{\bm{e}}_1,\\
&F_2 &= \bm{F}_{norm}\cdot \hat{\bm{e}}_2, \text{ and}\\
F_L = &F_3 &= \bm{F}_{norm}\cdot \hat{\bm{e}}_3.
\end{eqnarray}
Since the reported forces are without buoyancy effects, the $(1-\epsilon_s)$ term must be considered accordingly for both drag and lift while performing Euler-Lagrangian simulations. Due to the influence of neighbouring particles, the lateral force $F_2$ for each individual particle may not be equal zero, as shown in figure \ref{fig:force_independence} ($\Rey=100$ and $\epsilon_s=0.3$). However, due to symmetry, the average $F_2$ does equal zero. Therefore, $F_2$ is not considered in our further discussion. The torques about the above defined axes are
\begin{eqnarray}
&T_1 &= \bm{T}_{norm}\cdot \hat{\bm{e}}_1,\\
T_P = &T_2 &= \bm{T}_{norm}\cdot \hat{\bm{e}}_2, \text{ and}\\
&T_3 &= \bm{T}_{norm}\cdot \hat{\bm{e}}_3.
\end{eqnarray}
Here $T_P$ is the pitching torque acting on a particle. We show the three different torques for a flow through a random particle configuration at $\Rey=100$ and $\epsilon_s=0.3$  in figure \ref{fig:torque_Indpendence}. It can be observed that $T_1$ and $T_3$, though having some non-zero values, are statistically zero on average due to symmetry. The non-zero values are primarily due to hydrodynamic interactions with other particles. Only the average pitching torque $T_P$ (or $T_2$) remains non-zero for different $\phi$ and varies as $\sin\phi\cos\phi$. Though individual particles experience non-zero $T_1$ and $T_3$, they become zero at $\phi=0^\circ$ and $\phi=90^\circ$ respectively, where the axis of symmetry of the particle coincides with the measured axis for torque. This implies that the hydrodynamic interaction of particles does not induce a torque (or a spin) about the axis of symmetry of the particle.

\subsection{Validation}
\label{sec:multiparticle_validation}
\begin{figure}
\centering
\psfragfig[width=0.45\textwidth]{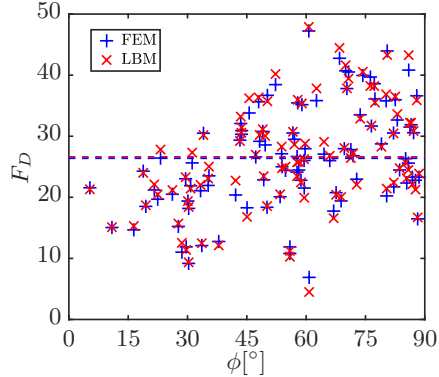}
\caption{$F_D$ obtained for individual particles in a random configuration from the LBM solver against FEM solver for $Re=100$ and $\epsilon_s=0.3$. The dashed lines in respective colours indicate the domain averages from the respective solvers.}
\label{fig:comsol_vs_lb3d}
\end{figure}

Sufficient validation has been done for our LBM code in the past for flow around isolated particles \citep{sanjeevi2017orientational,sanjeevi2018drag}. For a multiparticle configuration, we have chosen flow around a random assembly of 100 particles at $\Rey=100$ and $\epsilon_s=0.3$ and measure the $F_D$ experienced by the individual particles. The LBM results are compared with results from COMSOL Multiphysics, a body-fitted, unstructured mesh based FEM solver. The simulated LBM domain is of size $360^3$. The volume equivalent sphere diameter is $d_{eq} = 64.4$ lattice cells. The superficial velocity $u_s$ is 0.0414 and the kinematic viscosity $\nu$ is 0.08/3 in lattice units. The FEM solver domain is made of 2.1 million elements. The resulting drag forces are shown in figure \ref{fig:comsol_vs_lb3d}. A good agreement between LBM and FEM results can be observed. The average $F_D$ experienced by all particles in LBM and FEM solvers are 26.6 and 26.4 respectively. Also a good match in $F_D$ values for individual particles at different $\phi$ can be observed.

%

\section{Tests of configuration independence}
\label{sec:configIndependence}

\begin{figure}
\psfragfig[width=\textwidth]{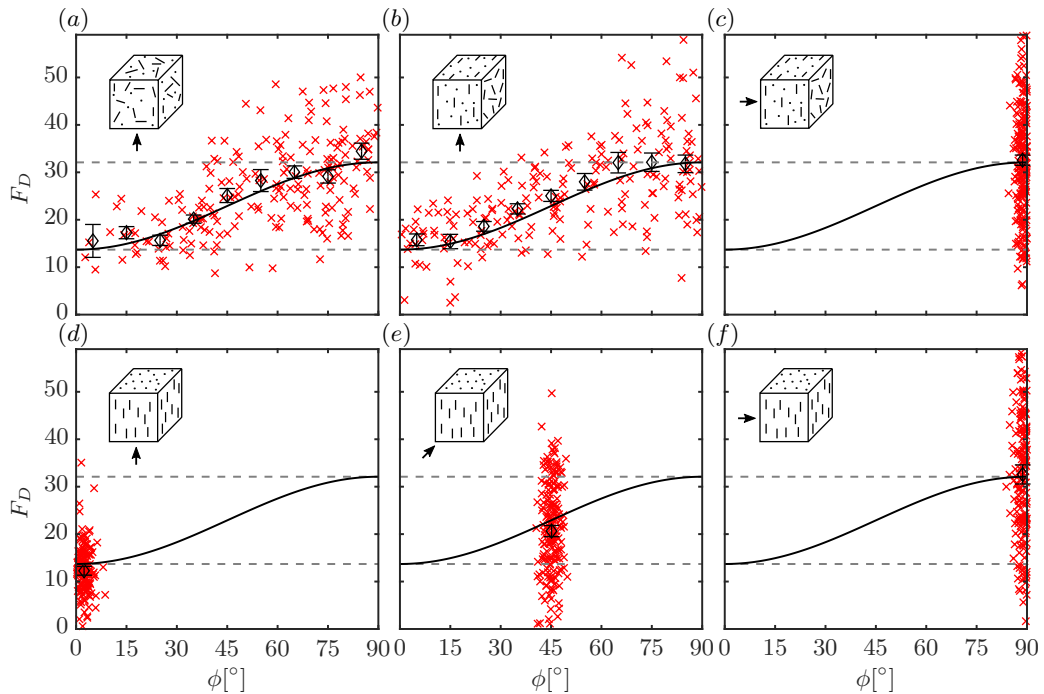}
\caption{Configuration independence phenomenon at $Re=100$ and $\epsilon_s=0.3$ for different configurations with different flow directions (arrow indicated).  $F_D$ distribution for different particles ($\cross$) with averages at regular $\phi$ intervals ($\Diamond$). ($a$) Random configuration, planar random configuration with flow ($b$) parallel and ($c$) perpendicular to the plane, unidirectional configuration with flow at ($d$) $0^\circ$, ($e$) $45^\circ$, and ($f$) $90^\circ$ with respect to the principal configuration director. The solid black line indicates the \emph{sine-squared} scaling.}
\label{fig:intermediateRe}
\end{figure}

Given a six-dimensional parameter space, exploring each dimension with approximately 5 simulations, results in a massive $5^6=15625$ simulations. Furthermore, closures must be created for drag, lift and torque as a function of this six-dimensional space. Before proceeding with these simulations, we tried to identify if there are any independent parameters specifically related to the mutual orientation of particles. In this section, we will show that the flow around a non-spherical particle assembly is independent of the mutual orientation of the particles themselves. This configuration independence removes the configuration parameters $S_1$, $S_2$ and flow angle parameters $\alpha$ and $\beta$ from the parameter space to be explored. We find that, when averaged over a number of particles, the only dependence that the particles exhibit regarding orientation is the particle's incident angle $\phi$ as in flow around single particles. Effectively, we will show that the flow problem depends only on the Reynolds number $\Rey$, solids volume fraction $\epsilon_s$ and the incident angle $\phi$ of individual particles with respect to the flow direction.

In the extremely dilute regimes, i.e. $\epsilon_s\rightarrow0$, it is already shown that there exists a sine-squared scaling of drag for elongated non-spherical particles \citep{sanjeevi2017orientational, sanjeevi2018drag}. In this section, we discuss the results of flow around different configurations at an intermediate solids volume fraction of $\epsilon_s=0.3$. Results of different configurations (in respective plot insets) at an intermediate $\Rey=100$ are shown in figure \ref{fig:intermediateRe} such as fully random, planar random with flows parallel and perpendicular to the planes and unidirectional configurations with principal directors at different angles. Though there exists scatter in the measured $F_D$, it can be observed that the average $F_D$ at different $\phi$ interval scales similar to sine-squared scaling as in our earlier works of isolated particles. In other words, the $F_D$ at any $\phi$ can be computed as
\begin{eqnarray}
F_{D,\phi} = F_{D,\phi=0^{\circ}}+(F_{D,\phi=90^{\circ}}-F_{D,\phi=0^{\circ}})\sin^2\phi .
\label{eq:FD_stokes}
\end{eqnarray}

\begin{figure}
\psfragfig[width=\textwidth]{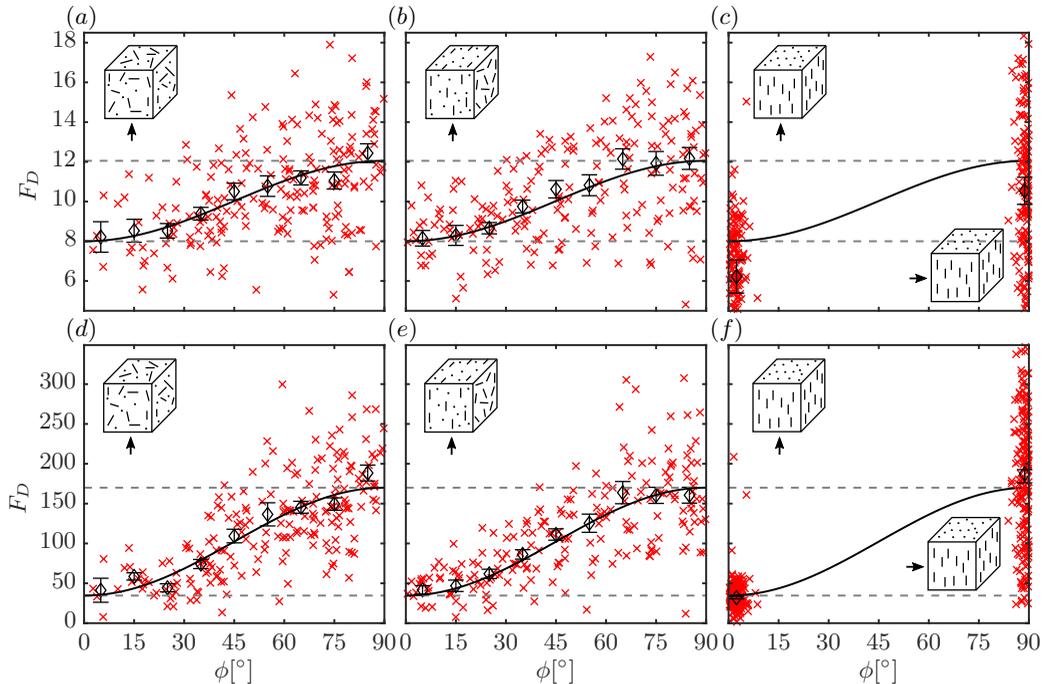}
\caption{Configuration independence phenomenon at moderate solids fraction $\epsilon_s=0.3$ for ($a-c$) $Re=0.1$ (low $\Rey$) and ($d-f$) $Re=1000$ (high $\Rey$) for different configurations and different flow directions (arrow indicated). $F_D$ distribution for different particles ($\cross$) with averages at regular $\phi$ intervals ($\Diamond$). ($a,d$) Random configuration, ($b,e$) planar random configuration with flow parallel to the plane, ($c,f$) combined results of unidirectional configuration with flow  $0^\circ$ and $90^\circ$ with respect to the principal configuration director. The solid black line indicates the respective \emph{sine-squared} scaling.}
\label{fig:lowHighRe}
\end{figure}

It is important to note that the \emph{same} values for $F_{D,\phi=0^\circ}$ and $F_{D,\phi=90^\circ}$ emerge for all configurations. Likewise, we also show that the scaling phenomenon extends to both Stokes and high $\Rey$ regimes in figure \ref{fig:lowHighRe}. With the sine-squared scaling behaviour (or the configuration independence) identified at $\epsilon_s=0$ and $\epsilon_s=0.3$, it can be inferred that the scaling is safely applicable in the region $0\leq\epsilon_s\leq0.3$. We have verified the same at $\epsilon_s=0.1$ and the results are not shown here for brevity. Though we observe the results are dependent on only 3 parameters, namely $\Rey$, $\epsilon_s$ and $\phi$, the simulation needs to be set up for only two parameters, namely $\Rey$ and $\epsilon_s$. With a sufficiently random configuration, the system involves different particle orientations covering all $\phi$. A caveat with a random configuration is that there are always very few particles near $\phi=0^\circ$, as shown in section \ref{sec:configurations}. Therefore, biased random configurations with more particles at $\phi=0^\circ$ are created and at least two simulations are performed for better statistics.

\begin{figure}
\psfragfig[width=\textwidth]{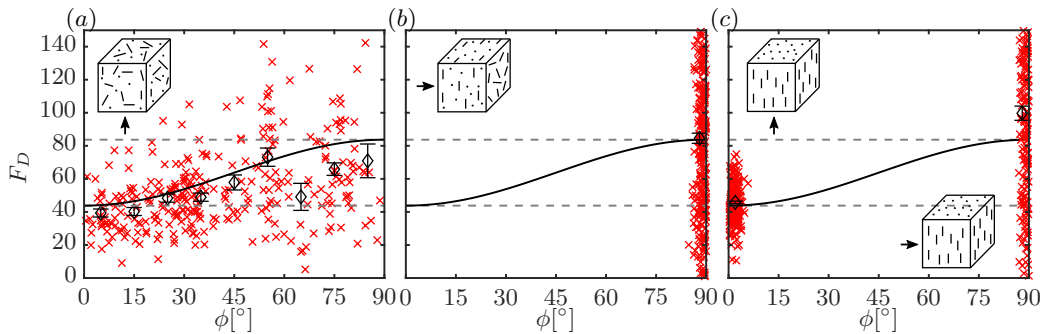}
\caption{Configuration independence phenomenon at dense solids fraction $\epsilon_s=0.5$ for $Re=100$ for different configurations and different flow directions (arrow indicated). $F_D$ distribution for different particles ($\cross$) with averages at regular $\phi$ intervals ($\Diamond$).}
\label{fig:densePackingRe100}
\end{figure}

We also observe the configuration independence phenomenon at $\epsilon_s=0.4$. The criterion considered to declare configuration independence phenomena is that the average drag results in a given $\phi$ range of different configurations are within 10\% deviation. In almost all cases, the deviations are within $\pm$ 5\%. However in a dense case with $\epsilon_s=0.5$, several more factors such as the mutual orientations, relative positions of particles, etc. influence the results. The $F_D$ distribution for such dense configurations at $\Rey=100$ and $\epsilon_s=0.5$ are given in figure \ref{fig:densePackingRe100}. Although these results can be predominantly parametrized by $\Rey$, $\epsilon_s$, and $\phi$, the influence of the additional parameters cannot be ignored. Therefore, specific cases of $\epsilon_s=0.5$ are performed with more simulations for better statistics.

\begin{figure}
\centering
\psfragfig[width=0.8\textwidth]{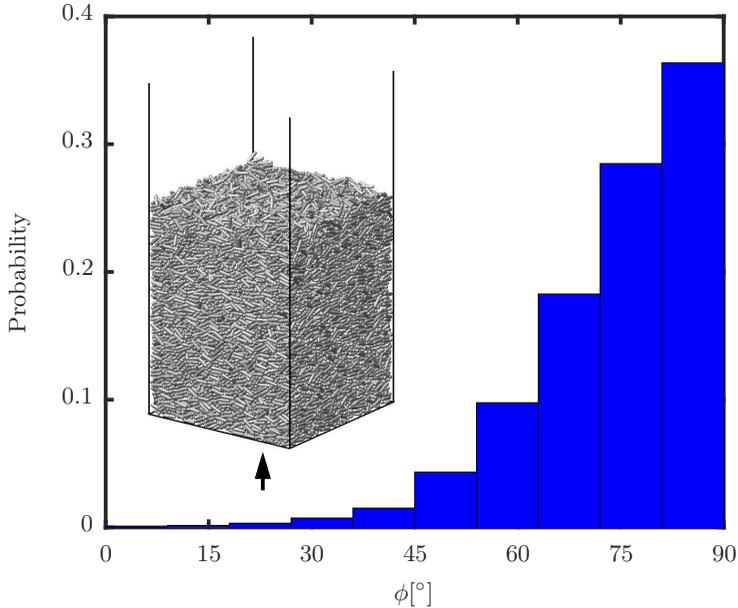}
\caption{Histogram of incident angle $\phi$ for a packed bed with 30000 particles. The arrow indicates the flow direction.}
\label{fig:packedBed}
\end{figure}

For a practical fluidization or other relevant gas-solid flow simulation, the densest configuration is most likely to occur when the particles are at bottom or at rest (e.g. before the start of fluidization). In such a dense condition, the particle configuration itself is dependent on the wall geometry. For a typical bed configuration with a flat wall at the bottom, the particles also roughly align in planes parallel to the bottom wall, i.e. a planar random configuration. \citet{pournin2005three} observed the same for particles poured freely from the top. Similarly, we also observe the same for a bed containing freely poured particles settled under gravity ($\epsilon_s=0.54$), as shown in figure \ref{fig:packedBed}. The bed contains 30000 particles and it can be observed that roughly 2/3 of all particles are in the range $\phi=70-90^\circ$ confirming our hypothesis. Given such criteria, the most relevant regime would be to generate an accurate fit for $F_{D,\phi=90^\circ}$ at high $\epsilon_s$. 

It should also be noted that with increasing aspect ratio of elongated particles, the maximum $\epsilon_s$ decreases for a packed bed \citep{williams2003random}. This is because the locking phenomenon is stronger with high aspect ratio particles. Unless the particles are packed with their orientations aligned, the decrease in peak $\epsilon_s$ for high aspect ratio elongated particles is unavoidable. Also, practical applications as shown in figure \ref{fig:packedBed} do not allow such long range ordering. A decreasing peak $\epsilon_s$ implies that the configuration independence phenomenon will be very applicable. With the observed sine-squared drag scaling, the pressure drop across a packed bed can be determined with the knowledge of the $\phi$ distribution alone.

In the subsequent sections, we will show that in the dilute and intermediate $\epsilon_s$ regimes, the influence of $\epsilon_s$ is nearly shape independent. This implies that the drag on isolated non-spherical particles can be combined with sphere-based multiparticle correlations for the voidage effect to mimic flow around assemblies of non-spherical particles upto intermediate $\epsilon_s$.

\subsection{Explored regimes}
\label{sec:explored_regimes}

\begin{figure}
\centering
\psfragfig[width=0.75\textwidth]{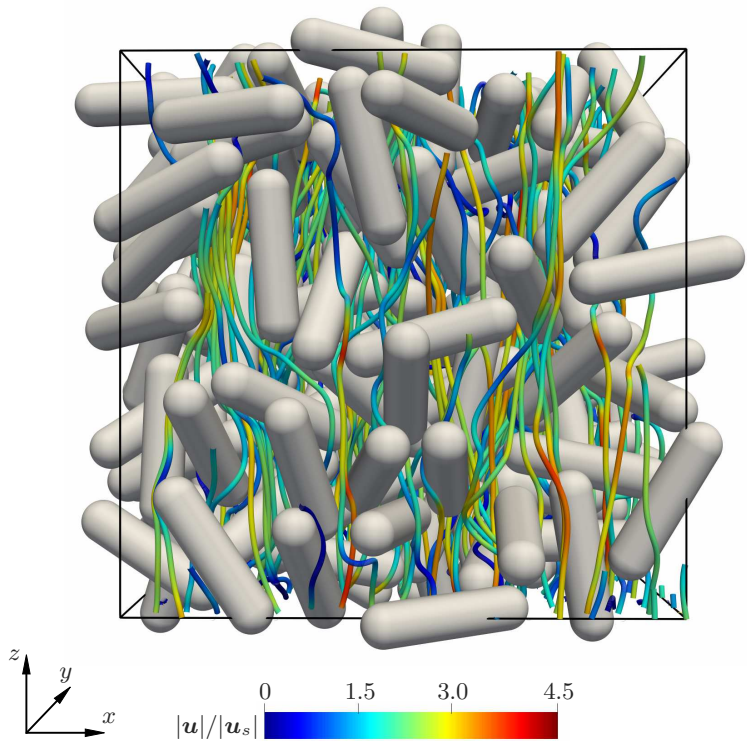}
\caption{Flow streamlines for a random configuration at $\Rey=100$ and $\epsilon_s=0.3$.}
\label{fig:stream_lines}
\end{figure}

\begin{figure}
\centering
\psfragfig[width=0.7\textwidth]{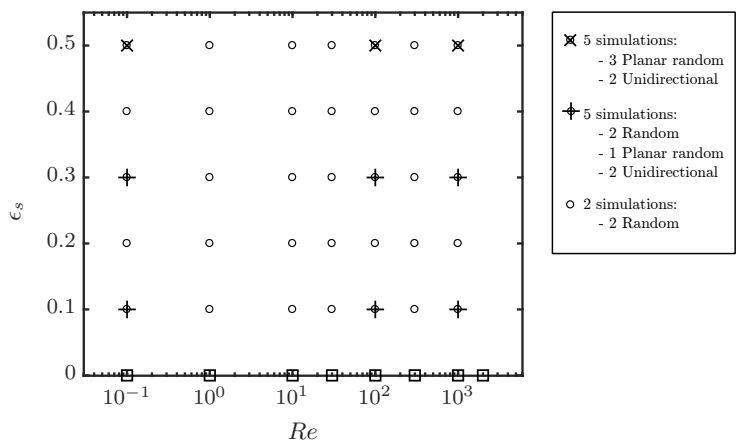}
\caption{Regime map containing the explored parameter space in the current work ($\circ$) and our previous work \citet{sanjeevi2017orientational} ($\square$). $+,\times$ indicate the regimes with extra simulations and tested for configuration independence.}
\label{fig:multiRegimeMap}
\end{figure}

In this section, we briefly explain the regimes explored in the current work and also explain the number of independent simulations performed per regime tested. An example of the flow stream lines for a random configuration at $\Rey=100$ and $\epsilon_s=0.3$ is shown in figure \ref{fig:stream_lines}. Until solids volume fractions of $\epsilon_s=0.35$, the generation of randomly orientation configurations is possible, as experienced by \citet{he2018variation} for prolate spheroids of aspect ratio 2.5. In our case, we are able to achieve random configurations upto $\epsilon_s=0.4$. However for denser configurations, it is difficult to generate a truly random configuration. For dense configurations of $\epsilon_s=0.5$, the particles have a natural tendency to orient to planar random or unidirectional orientation configurations. A truly random configuration with a finite number of particles, at such solids volume fraction, is not possible. This is due to a strong orientation bias imposed by neighbouring particles due to lack of inter-particle space. The explored regimes are indicated in figure \ref{fig:multiRegimeMap}. Overall, at least two simulations are performed for the explored regimes. However for specific cases of dilute and intermediate $\epsilon_s$, we performed 5 simulations with 2 random, 1 planar random with flow aligned to the plane and 2 unidirectional configurations with flow parallel and perpendicular to the principal director. For solids fraction $\epsilon_s=0.5$, 3 planar random configurations with flows aligned to the plane and 2 unidirectional configurations with flows parallel and perpendicular to the principal director are performed. For cases with more simulations, the results are accordingly weighted while making the fits.

\section{Results}
\subsection{Drag}
With sine-squared scaling valid as shown in section \ref{sec:configIndependence}, the drag experienced by a particle in a multiparticle system can be explained by the equation \ref{eq:FD_stokes} involving only the drag experienced at $\phi=0^\circ$ and $\phi=90^\circ$. Therefore, we propose to generate fits for $F_{D,\phi=0^\circ}$ and $F_{D,\phi=90^\circ}$ as a function of $\Rey$ and $\epsilon_s$ as
\begin{eqnarray}
\centering
&F_D(\Rey,\epsilon_s) &= F_{d,isol}\cdot(1-\epsilon_s)^2 + F_{\epsilon_s}+F_{\Rey,\epsilon_s}.
\label{eq:FD_fit}
\end{eqnarray}
The corresponding terms are as follows:
\begin{eqnarray}
&F_{d,isol}(\Rey) &= C_{d,isol}\frac{\Rey}{24},\\
&F_{\epsilon_s}(\epsilon_s) &= a\sqrt{\epsilon_s}(1-\epsilon_s)^2+\frac{b\epsilon_s}{(1-\epsilon_s)^2}, \text{ and }
\label{eq:Fd_epsilon_drag}\\
&F_{\Rey,\epsilon_s}(\Rey,\epsilon_s) &=  \Rey^c\epsilon_s^d \left(e(1-\epsilon_s)+\dfrac{f\epsilon_s^3}{(1-\epsilon_s)}\right)+g\epsilon_s(1-\epsilon_s)^2\Rey.
\label{eq:Fd_Re_drag}
\end{eqnarray}

Here, $C_{d,isol}$ is the isolated particle drag at given $\Rey$ as detailed in \citet{sanjeevi2018drag} for the considered particle (fibre) for both $\phi=0^\circ$ and $\phi=90^\circ$. The coefficients in equations \ref{eq:Fd_epsilon_drag} and \ref{eq:Fd_Re_drag} for both $F_{D,\phi=0^\circ}$ and $F_{D,\phi=90^\circ}$ are given in table \ref{tab:drag}. The average absolute deviation of the fits and simulation data are 3.5\% and 2\% for $F_{D,\phi=0^\circ}$ and $F_{D,\phi=90^\circ}$, respectively.

\begin{table}
\begin{center}
\def~{\hphantom{0}}
\begin{tabular}{clll}
\toprule
& \multicolumn{2}{c}{$F_D$}\\
\cline{2-3}
Coefficients    & $\phi=0^\circ$ & $\phi=90^\circ$ & $F_{L,mag}$ \\
\midrule
$a$ & 2              & 3               & 0.85        \\
$b$ & 11.3           & 17.2            & 5.4         \\
$c$ & 0.69           & 0.79            & 0.97        \\
$d$ & 0.77           & 3               & 0.75        \\
$e$ & 0.42           & 11.12           & -0.92       \\
$f$ & 4.84           & 11.12           & 2.66        \\
$g$ & 0              & 0.57            & 1.94       \\
\bottomrule
\end{tabular}
\caption{Coefficients of the fits for $F_D$ and $F_L$}
\label{tab:drag}
\end{center}
\end{table}

\begin{figure}
\centering
\psfragfig[width=\textwidth]{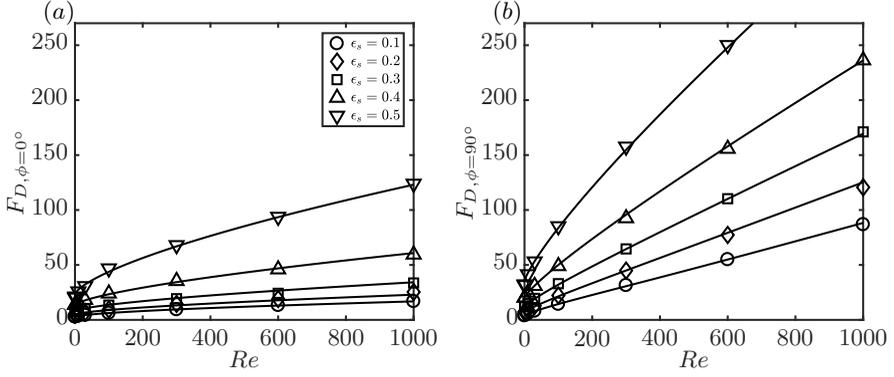}
\caption{The drag forces ($a$) $F_{D,\phi=0^\circ}$ and ($b$) $F_{D,\phi=90^\circ}$ at different $\Rey$ and $\epsilon_s$. The markers indicate simulation data and the solid lines are corresponding fits.}
\label{fig:Fd_Re_eps}
\end{figure}

\begin{figure}
\centering
\psfragfig[width=\textwidth]{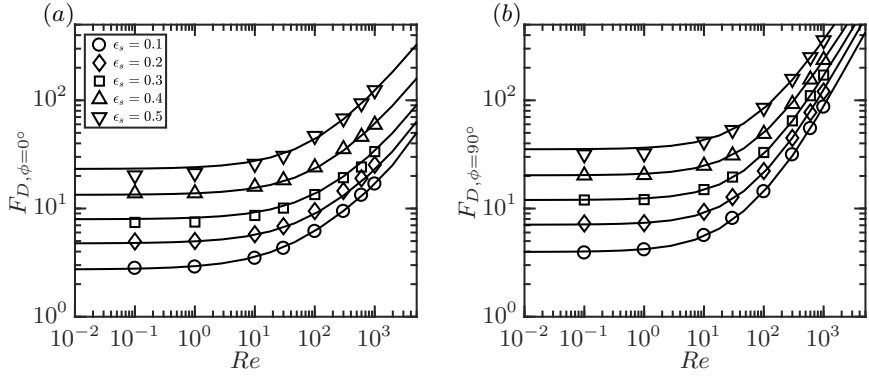}
\caption{The fits for ($a$) $F_{D,\phi=0^\circ}$ and ($b$) $F_{D,\phi=90^\circ}$ at different $\Rey$ and $\epsilon_s$ beyond the simulated regimes of $0.1\leq\Rey\leq1000$. The markers indicate simulation data and the solid lines denote corresponding fits.}
\label{fig:Fd_Re_eps_loglog}
\end{figure}

The simulated data and corresponding fits are shown in figure \ref{fig:Fd_Re_eps}. The fits follow the physical limits beyond the $\Rey$ range simulated as shown in figure \ref{fig:Fd_Re_eps_loglog}. In the Stokes flow limit, it can be observed that both $\phi=0^\circ$ and $\phi=90^\circ$ normalized drag becomes independent of $\Rey$. In the high $\Rey$ limit, the normalized drag approaches a linear dependency on $\Rey$.

\begin{figure}
\centering
\psfragfig[width=0.5\textwidth]{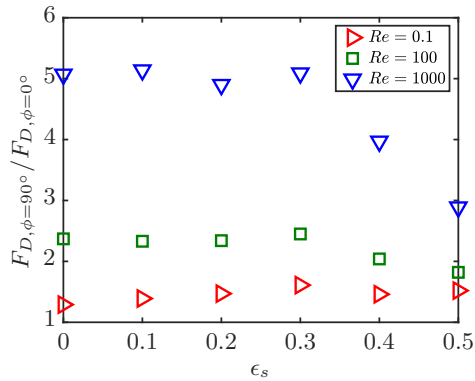}
\caption{Ratio of perpendicular to parallel drag $F_{D,\phi=90^\circ}/F_{D,\phi=0^\circ}$ from simulations for different $\Rey$ and $\epsilon_s$.}
\label{fig:DragRatios}
\end{figure}

\begin{figure}
\centering
\psfragfig[width=0.5\textwidth]{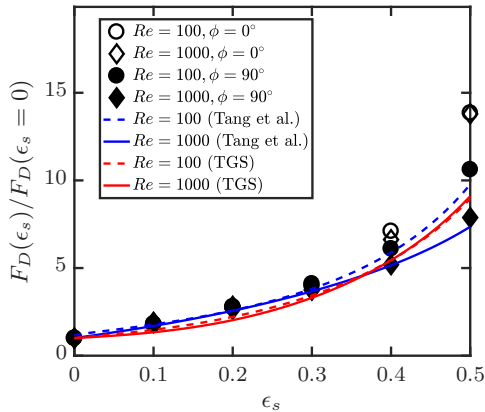}
\caption{Voidage effect on drag: $F_{D}(\epsilon_s)/F_{D}(\epsilon_s=0)$ for $\phi=0^\circ$ and $\phi=90^\circ$ in the inertial regimes as a function of $\epsilon_s$ for spherocylinders (this work, symbols), compared with voidage effect for spheres from literature.}
\label{fig:Fd_by_Fdisol_phicombined}
\end{figure}

\begin{figure}
\centering
\psfragfig[width=\textwidth]{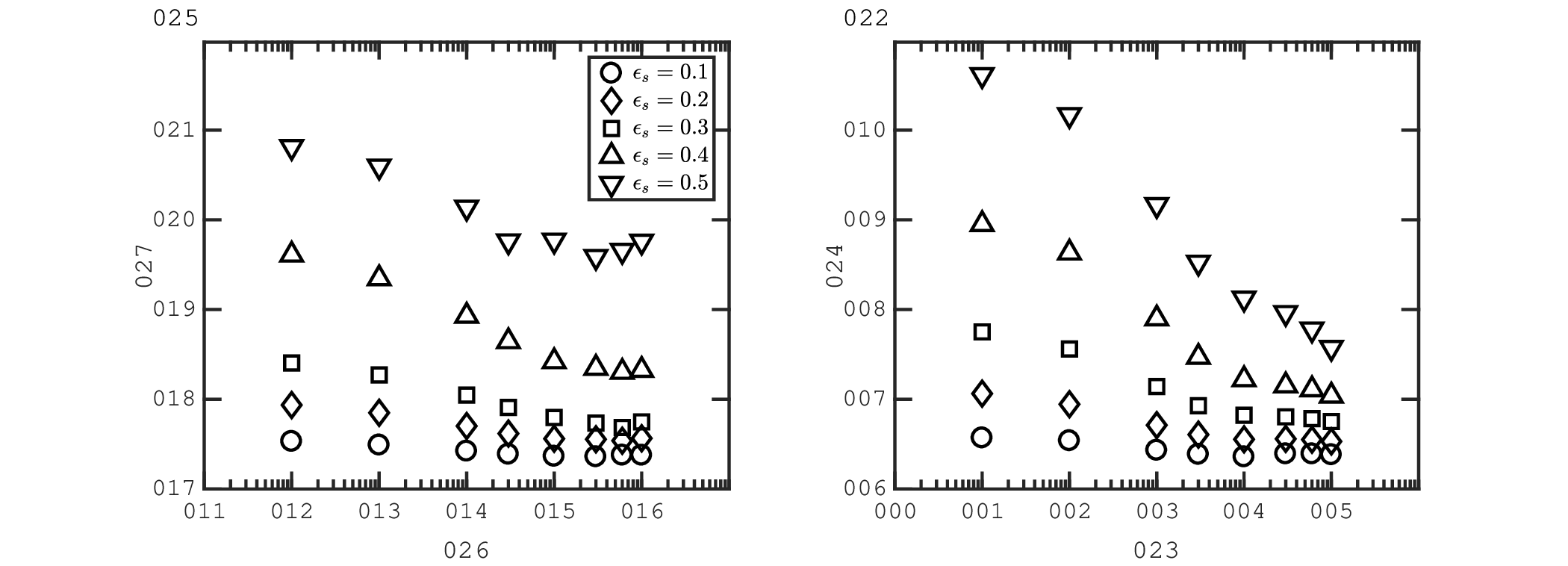}
\caption{$F_{D,\phi=0^\circ}(\epsilon_s)/F_{D,\phi=0^\circ}(\epsilon_s=0)$ and $F_{D,\phi=90^\circ}(\epsilon_s)/F_{D,\phi=90^\circ}(\epsilon_s=0)$ as a function of $\Rey$.}
\label{fig:Fd_by_Fdisol}
\end{figure}

The ratio of the perpendicular to parallel drag $F_{D,\phi=90^\circ}/F_{D,\phi=0^\circ}$ at different $\Rey$ and $\epsilon_s$ is shown in figure \ref{fig:DragRatios}. For low $\Rey$ ($\Rey=0.1$), the ratio remains constant at a value a little larger than 1 for all $\epsilon_s$. The reason for this is that at low $\Rey$, the particles experience stronger viscous effects. The viscous drag reduces and pressure drag increases with increasing $\phi$ at low $\Rey$. The same has been confirmed for isolated particles \citep{sanjeevi2017orientational} and for a multiparticle system \citep{he2018variation}. The combined viscous and pressure drag components result in a drag ratio close to 1 for the considered spherocylinders at low $\Rey$. Due to inertial dominance at moderate and large $\Rey$ ($\Rey\geq100$) we can observe a near constant drag ratio for solids volume fractions upto $\epsilon_s=0.3$ and a decrease in the ratio for $\epsilon_s>0.3$. Further, figure \ref{fig:DragRatios} gives an indication that for very dense crowding, i.e. at $\epsilon_s>0.5$, there is a possibility that $F_{D,\phi=90^\circ}/F_{D,\phi=0^\circ}$ tends back to approximately 1. Up to moderate crowding, although the flow is disturbed due to the presence of neighbouring particles, there is sufficient inter-particle space for flow to achieve uniformity. However with increased particle crowding, there appear pronounced fluctuations in flow velocities (see also section \ref{sec:flow_histograms}), resulting in a reduced drag ratio at high $\epsilon_s$. This is an important finding because the traditional approach of Euler-Lagrangian simulations involve combining isolated non-spherical particle drag with the voidage effects based on sphere assemblies. This would result in a constant drag ratio for all $\epsilon_s$. This in turn could affect Euler-Lagrangian simulation results, especially in predicting the minimum fluidization velocity as there exists a dense packing of particles. This mandates the need for the current work.

Figure \ref{fig:Fd_by_Fdisol_phicombined} shows a similar interesting observation: The scaling of the voidage effect $F_D(\epsilon_s) / F_D(\epsilon_s = 0)$ in the inertial regime (high $\Rey$ limit) is shape and orientation independent for $\epsilon_s\leq0.3$. Here, we have normalized the drag with respective isolated particle drag for different $\Rey$ and $\phi$. It can be observed that all the normalized points fall on a same trend until $\epsilon_s=0.3$. Similar normalized $F_D$ for spheres from \citet{yali2015new} at $\Rey=100$ and $\Rey=1000$ also show the same trend until $\epsilon_s=0.3$. Here, we use the isolated sphere drag correlation of \citet{Schiller1935drag} for the normalization. The predictions of \citet{tenneti2011drag} for spheres do not follow the exact trend for the voidage effects as observed from figure \ref{fig:Fd_by_Fdisol_phicombined}. It should be noted that \citet{tenneti2011drag} explored only until $\Rey=300$ in their work and extrapolation to high $\Rey$ may not apply. Therefore, the above discussion indicates that spherical drag correlations for the voidage effect, combined with isolated non-spherical particle drag correlations can be applied to dilute suspension simulations of non-spherical particles in the inertial regimes. For a given non-spherical particle, the effect of crowding ($\epsilon_s$) on $F_D$ is different for different $Re$ and $\phi$. Figure \ref{fig:Fd_by_Fdisol} shows the voidage effect ($F_D$ normalized by the corresponding isolated particle drag)  as a function of $\Rey$. It can be seen at low $\Rey$, the increase in drag due to crowding is comparable for both $\phi=0^\circ$ and $\phi=90^\circ$  at different $\epsilon_s$. At high $\Rey$, the increase in drag due to crowding with increasing $\epsilon_s$ is much stronger for $\phi=0^\circ$ compared to $\phi=90^\circ$. This also explains further the reason for the observed reduction in perpendicular to parallel drag ratios with increasing $\epsilon_s$ in figure \ref{fig:DragRatios}.

\begin{figure}
\centering
\psfragfig[width=0.5\textwidth]{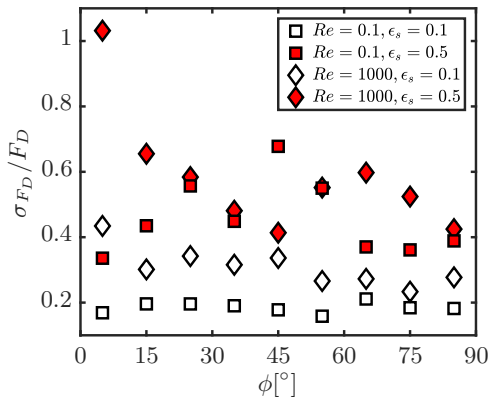}
\caption{The standard deviations $\sigma_{F_D}$ of the distribution of individual drag values, normalized by the corresponding average $F_D$ in different intervals of incident angle $\phi$.  Open symbols correspond to dilute configurations and filled symbols correspond to dense configurations.}
\label{fig:StdDev_FD}
\end{figure}

In the previous sections, we discussed the $F_D$ averaged over all particles with similar $\phi$. However, the distribution of $F_D$ within a $\phi$ interval is itself also a function of both $\Rey$ and $\epsilon_s$. The standard deviations of the distribution of drag measurements, normalized by the average $F_D$ in the corresponding interval, are plotted in figure \ref{fig:StdDev_FD}. It is important that the standard deviations are normalized by the respective average $F_D$, rather than against a single value, say $F_{D,\phi}=90^\circ$, for a given $\Rey$ and $\epsilon_s$. This is because with increasing $\Rey$, the ratio $F_{D,\phi=90^\circ}/F_{D,\phi=0^\circ}$ increases, as shown in figure \ref{fig:DragRatios} earlier. Therefore, using $F_{D,\phi=90^\circ}$ for normalization will make the standard deviations at $\phi=0^\circ$ appear insignificant at large $\Rey$.

For dilute configurations ($\epsilon_s=0.1$), we clearly observe that increasing $\Rey$ results in an increased $\sigma_{F_D}/F_D$ at all $\phi$. It should be noted that the absolute magnitudes of $F_D$ at $\Rey=1000$ are much larger than at $\Rey=0.1$. Despite the normalization by these larger values, we observe increased standard deviations for higher $\Rey$. This is because at low $\Rey$, the viscous effects dominate, resulting in long-range flow uniformity. Conversely, at high $\Rey$, the boundary layers are thinner and flow wakes are stronger. This results in high non-uniformity in the incoming flow on each particle, and thereby large fluctuations in the hydrodynamic forces. For dense particle configurations ($\epsilon_s=0.5$), it can be observed that $\sigma_{F_D}/F_D$ increases relative to dilute conditions, with a higher standard deviation for higher $\Rey$. The reason for higher spread in $F_D$ is due to the fact the particles locally encounter highly non-uniform incoming flows when there is more crowding.

\subsubsection{Comparison with other literature}
\begin{figure}
\centering
\psfragfig[width=\textwidth]{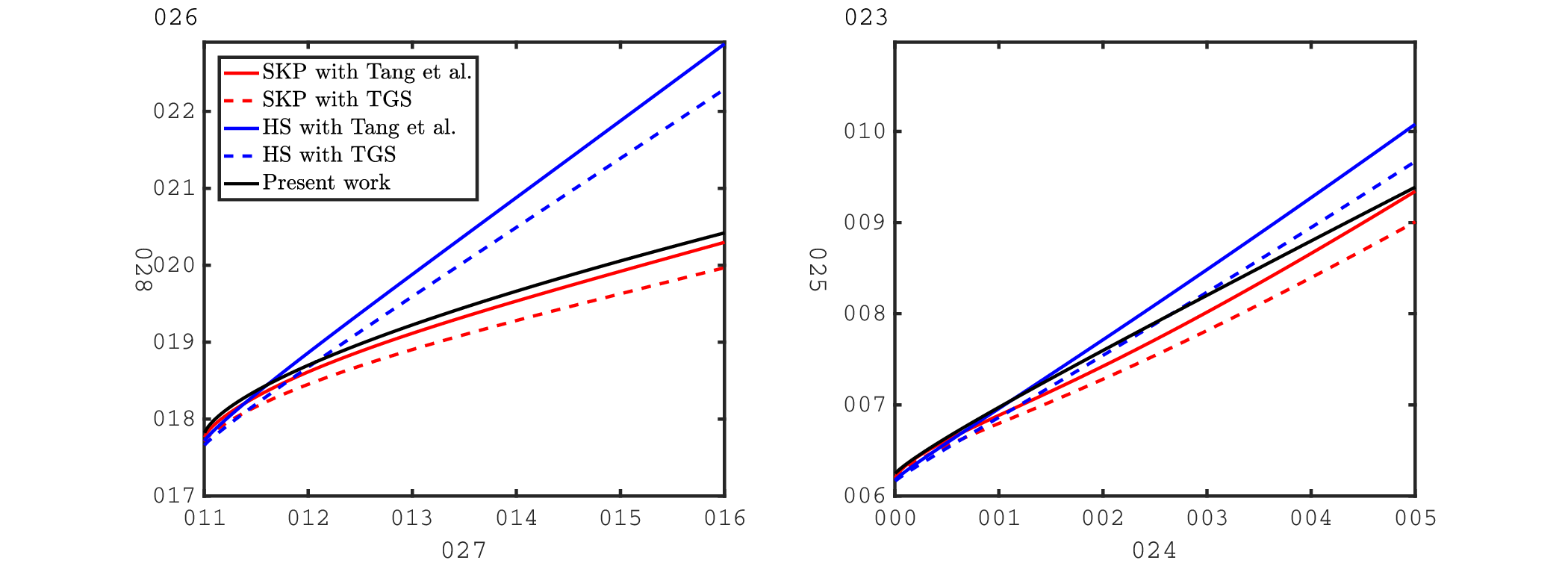}
\caption{Comparison of $F_D$ for ($a$) $\phi=0^\circ$ and ($b$) $\phi=90^\circ$ for $\epsilon_s=0.3$. SKP denotes \citet{sanjeevi2018drag}, HS denotes \citet{holzer2008new}, and TGS denotes \citet{tenneti2011drag}. The solid black line is equation \ref{eq:FD_fit}.}
\label{fig:FD_multi_eps0_3}
\end{figure}

\begin{figure}
\centering
\psfragfig[width=\textwidth]{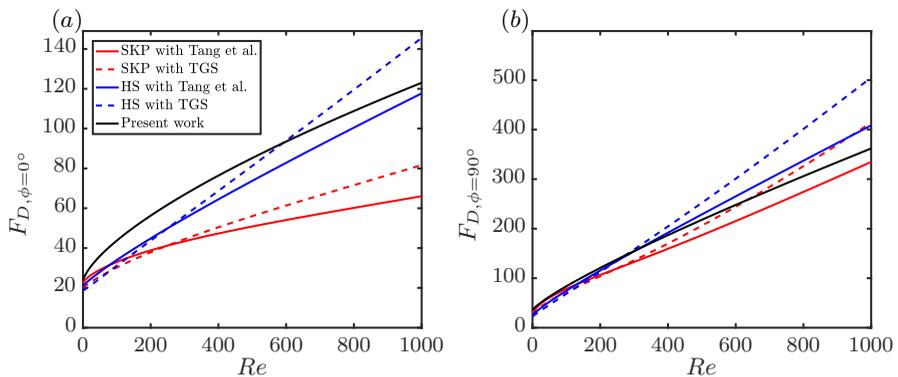}
\caption{Comparison of $F_D$ for ($a$) $\phi=0^\circ$ and ($b$) $\phi=90^\circ$ for $\epsilon_s=0.5$. SKP denotes \citet{sanjeevi2018drag}, HS denotes \citet{holzer2008new}, and TGS denotes \citet{tenneti2011drag}. The solid black line is equation \ref{eq:FD_fit}.}
\label{fig:FD_multi_eps0_5}
\end{figure}

Given the unavailability of multiparticle correlations for non-spherical particles, we combine the available literature results on isolated non-spherical particles with voidage effects based on spheres. For this, we normalize the multiparticle drag of spheres with the isolated sphere \citet{Schiller1935drag} correlation and multiply with the isolated non-spherical particle drag. The results are shown in figures \ref{fig:FD_multi_eps0_3} and \ref{fig:FD_multi_eps0_5} for $\epsilon_s=0.3$ and $\epsilon_s=0.5$, respectively. The isolated particles drag law used are SKP \citep{sanjeevi2018drag} and HS \citep{holzer2008new}. They are accordingly combined with the multiparticle effects of TGS \citep{tenneti2011drag} and \citet{yali2015new} for spheres. In the moderately crowded regime ($\epsilon_s=0.3$), our earlier suggestion of combining isolated non-spherical particle drag with multiparticle effects from spheres works well. For example, the combination of SKP with \citet{yali2015new} follows nearly the same trend as that of the current work (equation \ref{eq:FD_fit}). This can be observed for both $\phi=0^\circ$ and $\phi=90^\circ$. However for dense regimes ($\epsilon_s=0.5$), it can be observed that the combination of SKP with \citet{yali2015new} does not agree well with the present work for $\phi=0^\circ$. At the same time, the combination with the HS \citep{holzer2008new} isolated drag law seem to be closer to the current work for $\epsilon_s=0.5$. Such an agreement must be considered with care. The decent agreement occurs because HS possesses high drag values for $\phi=0^\circ$ (for $\epsilon_s=0$), in combination with a weak voidage effect for spheres. On the other hand, SKP with TGS or \citet{yali2015new} show decent agreement with the present work for $\phi=90^\circ$.

\subsection{Lift}
\begin{figure}
\centering
\psfragfig[width=\textwidth]{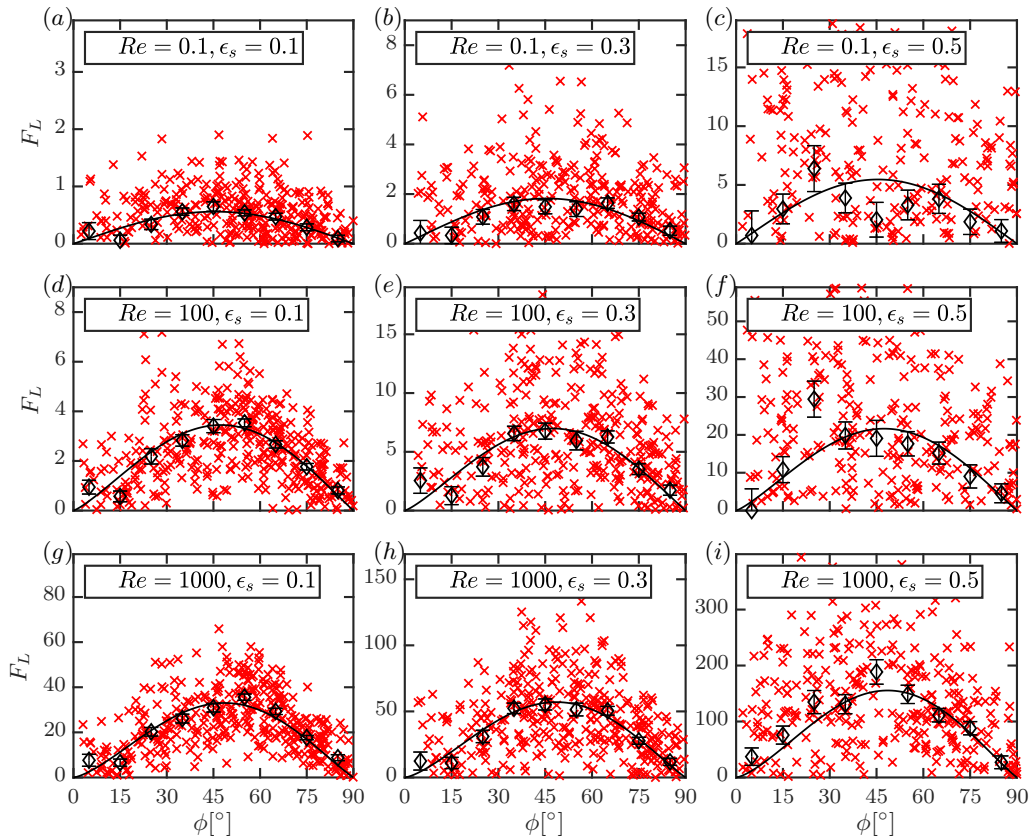}
\caption{Distributions of normalized lift forces $F_L$ ($\cross$) with averages at regular $\phi$ intervals ($\Diamond$) for different $\Rey$ and $\epsilon_s$. The solid line denotes the $F_{L,\phi}$ fit (equation \ref{eq:FL_RePhi}). Each plot includes data from two independent simulations with a total 400 data points. It should be noted that the scales are different for each plot.}
\label{fig:FLDist}
\end{figure}
The normalized lift $F_{l,\phi}$ on a single elongated particle from \citet{sanjeevi2018drag} is given by
\begin{eqnarray}
&F_{l,\phi}(\Rey,\phi) &= F_{l,isol}\cdot S_{f,\phi}, \text{ with }\\
\label{eq:cl_isol}
&F_{l,isol}(\Rey) &= \left(\frac{b_1}{Re}+\frac{b_2}{Re^{b_3}}+\frac{b_4}{Re^{b_5}}\right)\frac{Re}{24}, \text{ and }\\
&S_{f,\phi}(\Rey,\phi)&=\sin\phi^{(1+b_6 Re^{b_7})}\cos\phi^{(1+b_{8} Re^{b_{9}})}.
\end{eqnarray}
Here, $S_{f,\phi}$ is the scaling function dependent on $\Rey$ and $\phi$. The coefficients $b_i$ are accordingly listed in the mentioned literature. In particular, the coefficients $b_6$ to $b_9$ describe the amount of skewness of the lift coefficient on a single elongated particle around $\phi = 45^\circ$. In the current work, we observe the same skewness for the multiparticle system at different $\Rey$. Therefore, we assume the term $S_{f,\phi}$ remains the same for the multiparticle system. The normalized lift $F_L$ for a multiparticle system takes the following form:
\begin{eqnarray}
&F_{L,\phi}(\Rey,\epsilon_s,\phi) &= F_{L,mag}(\Rey,\epsilon_s)\cdot S_{f,\phi}(\Rey,\phi).
\label{eq:FL_RePhi}
\end{eqnarray}
The functional form of $F_{L,mag}(\Rey,\epsilon_s)$ remains similar to that of the drag and is given by
\begin{eqnarray}
\centering
&F_{L,mag}(\Rey,\epsilon_s) &= F_{l,isol}(\Rey)\cdot(1-\epsilon_s)^2 + F_{\epsilon_s}(\epsilon_s)+F_{\Rey,\epsilon_s}(\Rey,\epsilon_s)
\label{eq:FL_fit}
\end{eqnarray}
with
\begin{eqnarray}
&F_{\epsilon_s}(\epsilon_s) &= a\sqrt{\epsilon_s}(1-\epsilon_s)^2+\frac{b\epsilon_s}{(1-\epsilon_s)^2}, \text{ and }\\
&F_{\Rey,\epsilon_s}(\Rey,\epsilon_s) &=  \Rey^c\epsilon_s^d \left(e(1-\epsilon_s)+\frac{f\epsilon_s^3}{(1-\epsilon_s)}\right)+g\epsilon_s(1-\epsilon_s)^2\Rey.
\end{eqnarray}
The corresponding coefficients are given in table \ref{tab:drag}. The proposed lift correlation has around 5\%  average absolute deviation with respect to the simulation results. The comparison of the $F_L$ from simulations and the proposed correlation is shown in figure \ref{fig:FLDist}.

\subsubsection{A simplified lift function}
\begin{figure}
\centering
\psfragfig[width=0.5\textwidth]{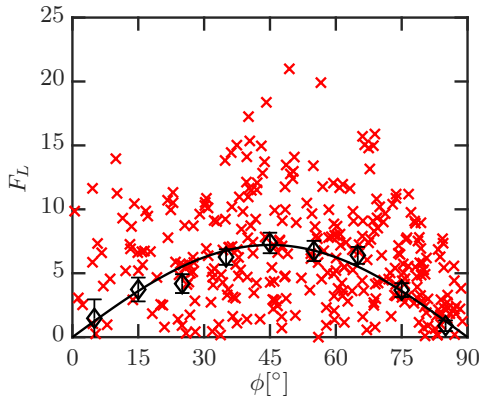}
\caption{Distribution of $F_L$ ($\cross$) for $\Rey=100$ and $\epsilon_s=0.3$ with averages ($\Diamond$) in regular $\phi$ intervals. The solid black line indicates the corresponding simple fit based on equation \ref{eq:FL_stokes_hypothesis}. The fit includes data from two different simulations totalling 400 data points.}
\label{fig:flHypExample}
\end{figure}

\begin{figure}
\centering
\psfragfig[width=0.5\textwidth]{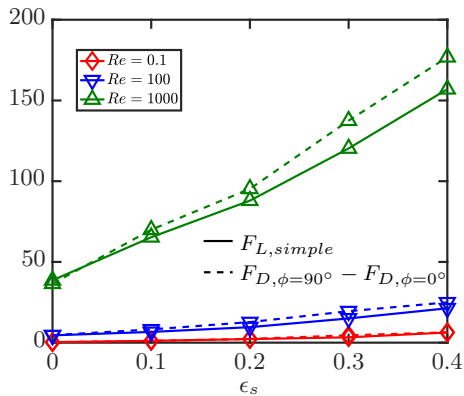}
\caption{Comparison of $F_{D,\phi=90^\circ} - F_{D,\phi=0^\circ}$ with $F_{L,simple}$ at different $\Rey$ and $\epsilon_s$. The difference $F_{D,\phi=90^\circ} - F_{D,\phi=0^\circ}$ is based on simulation data itself and not on the corresponding $F_D$ fits.}
\label{fig:FlmaxVsFdDiff}
\end{figure}

In our earlier works \citep{sanjeevi2017orientational, sanjeevi2018drag}, we have shown successfully that for elongated particles, the relation between lift and drag in the Stokes flow regime can be successfully used for higher $\Rey$ flows too. In other words, $F_L$ at different $\phi$ can be computed as
\begin{eqnarray}
F_{L,\phi}=(F_{D,\phi=90^{\circ}}-F_{D,\phi=0^{\circ}})\sin\phi \cos\phi.
\label{eq:FL_stokes}
\end{eqnarray}
In this section, we show that equation \ref{eq:FL_stokes} is a reasonable approximation even for a multiparticle system. This implies that the scaling law is valid not only just for different $\Rey$ but even for different $\epsilon_s$. Given a measured $F_L$ distribution from simulations at a given $\Rey$ and $\epsilon_s$, the data can be fitted in a simple form as
\begin{eqnarray}
F_{L,\phi}=F_{L,simple}\sin\phi \cos\phi.
\label{eq:FL_stokes_hypothesis}
\end{eqnarray}
Here, $F_{L,simple}$ is a fit parameter that best describes the simulation data. An example for such a fit for $\Rey=100$ and $\epsilon_s=0.3$ is given in figure \ref{fig:flHypExample}. The comparison of the Stokes regime lift law (equation \ref{eq:FL_stokes}) and our hypothesis (equation \ref{eq:FL_stokes_hypothesis}) is shown in figure \ref{fig:FlmaxVsFdDiff} and it can be observed that there is a good agreement. The highest absolute deviation observed between the equations is still less than 20\% and average absolute deviation is around 12\%.  Therefore in Euler-Lagrangian simulations, in the absence of explicit lift data, equation \ref{eq:FL_stokes} can be applied to include the effects of lift with acceptable accuracy. This implies that in the often-used approach of using \citet{holzer2008new} type drag correlations, combined with sphere-based voidage effect correlations in Euler-Lagrangian simulations, one can also include lift effects based on equation \ref{eq:FL_stokes}. In the following section, we will show the importance of including lift, as it is often of comparable magnitude to drag at high $\Rey$.

\subsubsection{Importance of lift compared to drag}
\begin{figure}
\centering
\psfragfig[width=0.8\textwidth]{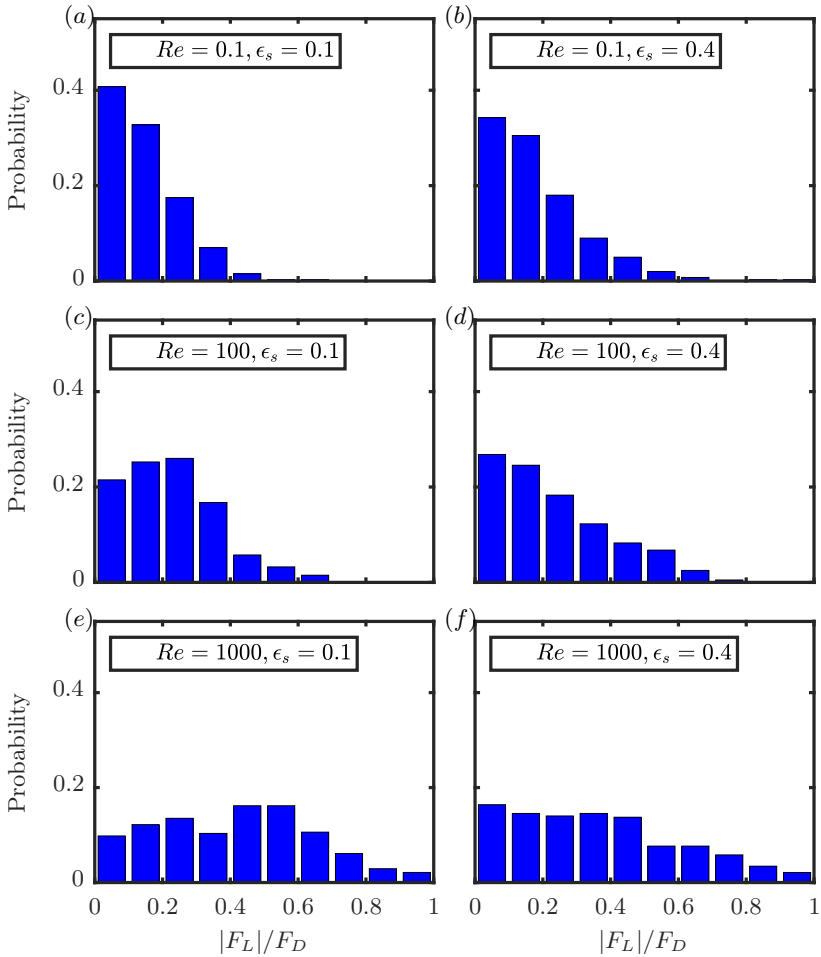}
\caption{Distribution of lift force on individual particles normalized by corresponding drag force on each particle at different $\Rey$ and $\epsilon_s$.}
\label{fig:liftDragRatios}
\end{figure}
In Euler-Lagrangian simulations, the effect of lift forces is often neglected. This is because there is not much literature on non-spherical particle lift correlations. In this section, we analyse the magnitudes of lift compared to the drag on individual non-spherical particles at different $\Rey$ and $\epsilon_s$. Figure \ref{fig:liftDragRatios} shows the distributions of the magnitude of the lift force relative to the drag force on each particle  $|F_L|/F_D$. It can be observed that for Stokes flow ($\Rey=0.1$), most particles experience lift which is about one order of magnitude smaller than the drag. However for high $\Rey$ ($\Rey=1000$), the distribution is much more wider spread and there are even some particles with $|F_L|/F_D=1$. This emphasizes the need for including lift in Euler-Lagrangian simulations, especially while handling Geldart D particles, where the encountered particle $\Rey$ is high. With increasing $\epsilon_s$, a different interesting observation can be made. In the low $\Rey$ regime, increasing $\epsilon_s$ results in an increased probability of particles experiencing high lift magnitudes compared to the drag. On the contrary, at high $\Rey$ ($\Rey=1000$), increasing $\epsilon_s$ results in the $|F_L|/F_D$ distribution skewing to the left. It should be noted that the highest $\epsilon_s$ shown in figure \ref{fig:liftDragRatios} is $\epsilon_s=0.4$ as opposed to $\epsilon_s=0.5$, the highest $\epsilon_s$ explored. This is because random configurations are not possible for $\epsilon_s=0.5$. To ensure consistency, all results shown in figure \ref{fig:liftDragRatios} are based on random configurations.

\subsection{Torque}
\begin{table}
\begin{center}
\begin{tabular}{cl}
\toprule
Coefficients    & $T_{P,mag}$  \\
\midrule
$a$ & 0.82            \\
$b$ & 1.44            \\
$c$ & 1.07            \\
$d$ & 5.48            \\
$e$ & 0.223           \\
\bottomrule
\end{tabular}
\caption{Coefficients of the fits for $T_{P,mag}$}
\label{tab:torque}
\end{center}
\end{table}

\begin{figure}
\centering
\psfragfig[width=\textwidth]{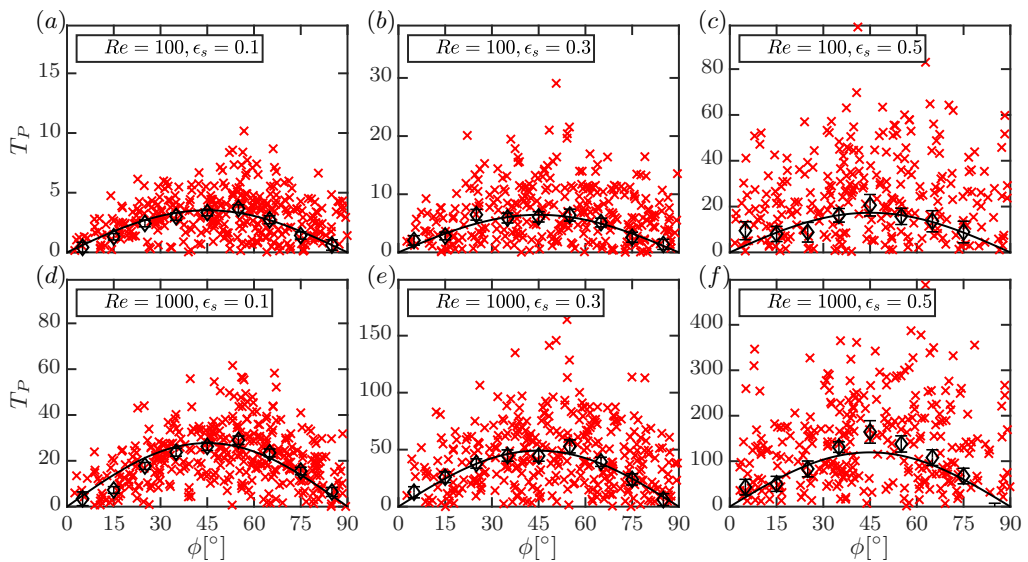}
\caption{Distributions of $T_P$ ($\cross$) with averages at regular $\phi$ intervals ($\Diamond$) for different $\Rey$ and $\epsilon_s$. The solid black line denotes $T_{P,\phi}$ fit (equation \ref{eq:Tp_stokes}). Each plot includes data from two independent simulations with each containing 400 data points. It should be noted that the scales are different for each plot.}
\label{fig:TpDist}
\end{figure}

For an isolated non-spherical particle, the torque correlation \citep{sanjeevi2018drag} is given by:

\begin{eqnarray}
&T_{p,\phi}(\Rey,\phi) &= T_{p,isol}(\Rey)\cdot S_{\phi}(\Rey,\phi), \text{ with }\\
\label{eq:ct_isol}
&T_{p,isol}(\Rey) &= \left(\frac{c_1}{Re^{c_2}}+\frac{c_3}{Re^{c_4}}\right)\frac{Re}{32}, \text{ and }\\
&S_{\phi}(\Rey,\phi)&=\sin\phi^{(1+c_5 Re^{c_6})}\cos\phi^{(1+c_{7} Re^{c_{8}})}.
\end{eqnarray}

The $\Rey$ dependent skewness terms $c_5,c_6,c_7,c_8$ equal zero for an isolated spherocylinder resulting in a symmetric distribution for $\phi$ around $45^\circ$. Likewise, we also observe a near symmetric distribution of torque at different $\Rey$ and $\epsilon_s$ for the multiparticle configuration (see figure \ref{fig:TpDist}). Unlike drag and lift, for an isolated non-spherical particle, the pitching torque vanishes for all $\phi$ in the Stokes flow regime. We observe the same for the multiparticle configuration. Therefore, the proposed correlation for the torque $T_P$ is applicable only in the inertial regime ($10<\Rey\leq1000$) and is given by
\begin{eqnarray}
&T_{P,\phi}(\Rey,\epsilon_s,\phi)&=T_{P,mag}(\Rey, \epsilon_s)\cdot\sin\phi \cos\phi, \text{ with }
\label{eq:Tp_stokes} \\
&T_{P,mag}(\Rey,\epsilon_s) &= T_{p,isol}(\Rey)\cdot(1-\epsilon_s)^2 + T_{\Rey,\epsilon_s}(\Rey,\epsilon_s).
\label{eq:TP_fit}
\end{eqnarray}
The corresponding terms in the scaling are as follows:
\begin{eqnarray}
T_{\Rey,\epsilon_s}(\Rey,\epsilon_s) &=  \Rey^a\epsilon_s^b \left(c(1-\epsilon_s)+\dfrac{d\epsilon_s^3}{(1-\epsilon_s)}\right)+e\epsilon_s(1-\epsilon_s)^2\Rey.
\end{eqnarray}

\begin{figure}
\centering
\psfragfig[width=0.5\textwidth]{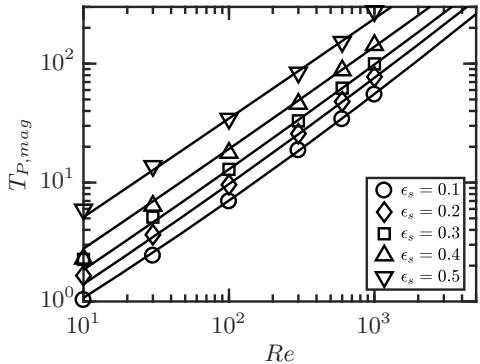}
\caption{$T_{P,mag}$ at different $\Rey$ and $\epsilon_s$. The markers indicate simulation data and the solid line denotes fit at corresponding $\epsilon_s$.}
\label{fig:TpFit}
\end{figure}

The average absolute deviation between equation \ref{eq:Tp_stokes} and corresponding simulation data is 3\%. It should be noted that $T_{P,mag}$ in equation \ref{eq:TP_fit} maps only the magnitude of the torque for different $\Rey$ and $\epsilon_s$. The $\phi$ dependence is included separately with the $sine$ and $cosine$ terms. The comparison of $T_{P,mag}$ and the corresponding simulation data are given in figure \ref{fig:TpFit}. Given a symmetric form for $T_{P,\phi}$, the $T_{P,mag}$ is equal to twice the magnitude of $T_{P,\phi=45^\circ}$ since $\sin\phi\cos\phi=1/2$ at $\phi=45^\circ$. From figure \ref{fig:TpFit}, it can be observed that $T_{P,mag}$ roughly follows the same power law dependence on $\Rey$ for different $\epsilon_s$ because the slopes are similar. This is in contrast to the drag trends in figure \ref{fig:Fd_Re_eps_loglog}, where the trend starts from zero slope at low $\Re$ to a constant slope at high $\Rey$. The reason is that the torque vanishes at low $\Rey$ for all $\epsilon_s$. The distributions of torque $T_P$ for different $\Rey$, $\epsilon_s$ and $\phi$ are given in figure \ref{fig:TpDist}.

\subsection{Flow histograms}
\label{sec:flow_histograms}
\begin{figure}
\centering
\psfragfig[width=0.8\textwidth]{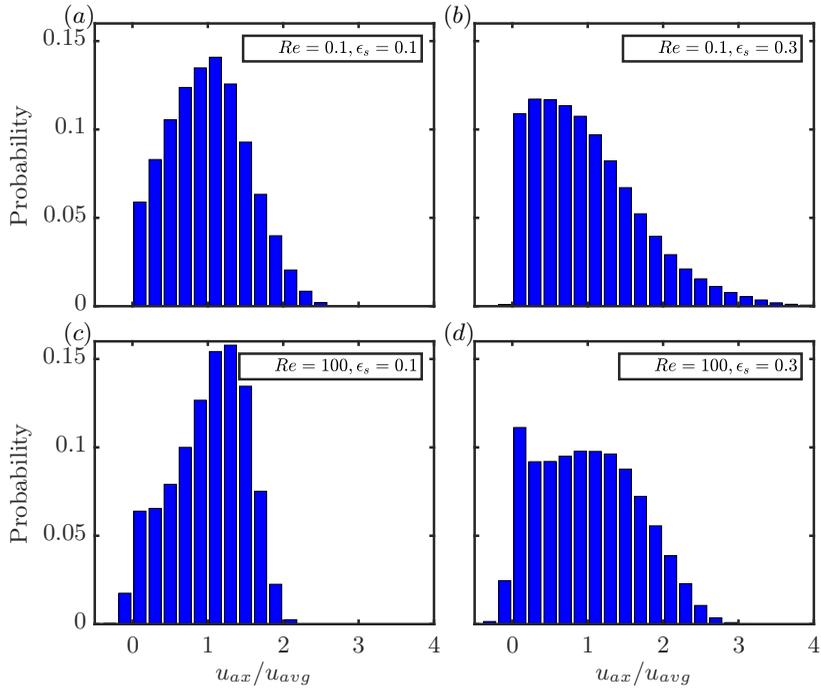}
\caption{Axial velocity distributions at different $Re$ and $\epsilon_s$ for a random configuration.}
\label{fig:velHist}
\end{figure}

\begin{figure}
\centering
\psfragfig[width=0.8\textwidth]{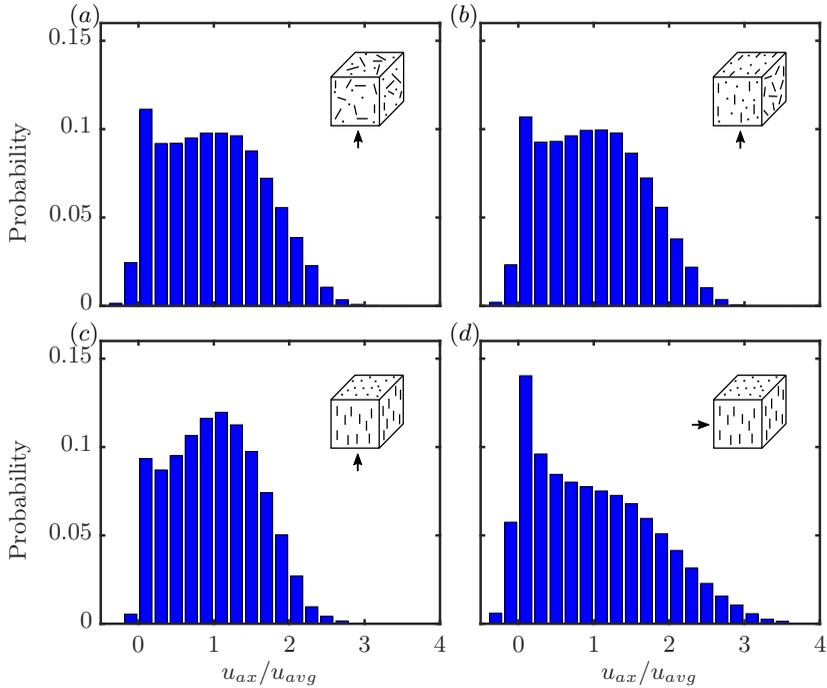}
\caption{Axial velocity distributions for different configurations at $\Rey=100$ and $\epsilon_s=0.3$.}
\label{fig:velConfigIndependence}
\end{figure}

In the previous sections, we discussed the influence of the flow on the hydrodynamic forces and torques on the particles. The flow around particulate assemblies can also be viewed as flow through a porous medium. In this section, we discuss the results of the influence of the particles on the flow distribution.

The probability distributions of the normalized axial flow velocities ($u_{ax}/u_{avg}$) at different $\Rey$ and $\epsilon_s$ for random configurations are given in figure \ref{fig:velHist}. Here, the normalization is done against the average axial velocity $u_{avg}= u_s/(1-\epsilon_s)$ rather than the superficial velocity $u_s$ to ensure a fair comparison for different $\epsilon_s$. Only the velocities of fluid cells are shown here and the zero velocities in the solid cells are ignored. It can be observed that with increasing $\Rey$, the spread of the velocity distribution becomes narrower. This can be simply attributed to the increased inertial effects and thinner boundary layers for increasing $\Rey$. Interestingly, the high $\Rey$ flows also demonstrate some negative velocities corresponding to wake effects. With increasing $\epsilon_s$, the peaks of the distribution shift towards the left and the distribution itself spreads wider. This implies that the increased presence of particle surfaces at higher $\epsilon_s$ pulls the velocities of fluid cells towards zero (hence the left skewness). At the same time, the fluid accelerates in the bulk regions further removed from the particle surfaces resulting in increased velocities (and hence a wider distribution) to maintain the desired $u_s$.

It is also interesting to investigate the velocity distributions for different configurations for a given $\Rey$ and $\epsilon_s$. The distributions of $u_{ax}/u_{avg}$ at $\Rey=100$ and $\epsilon_s=0.3$ for different configurations are plotted in figure \ref{fig:velConfigIndependence}. Given sufficient randomness of particles, as in random and planar random configurations (see figure \ref{fig:velConfigIndependence} ($a$) and ($b$)), the velocity distributions are nearly identical. However, velocity distributions can be different for different configurations, as can be observed for the unidirectional configurations with flow parallel and perpendicular to the principal director (see figure \ref{fig:velConfigIndependence} ($c$) and ($d$)). Among the different configurations shown, the unidirectional configuration with flow parallel to principal director has the least recirculation, as is evident from the least number of fluid cells with negative velocities ($u_{ax}/u_{avg}<0$). At the same time, as expected, the unidirectional configuration with flow perpendicular to principal director has the highest amount of recirculation. Overall, we can infer that there is no dependency between the configuration independence phenomenon, explained in section \ref{sec:configIndependence}, and the flow velocity distribution of different configurations. The variation in forces at different incident angles $\phi$ is mainly arising from the pressure forces. The same can also be confirmed from the multiparticle work of \citet{he2018variation}, which is also in line with our finding for isolated non-spherical particles \citep{sanjeevi2017orientational}.

\section{Conclusion}
The flow around assemblies of axisymmetric, non-spherical particles has been studied extensively using the multi-relaxation-time lattice Boltzmann method. The performed simulations are from the Stokes flow regime to high $\Rey$ ($0.1\leq\Rey\leq1000$) at different solids volume fraction $\epsilon_s$ ($\epsilon_s\leq0.5$) and different mutual orientations of particles.

In general, forces on random assemblies of spheres are only dependent on $\Rey$ and $\epsilon_s$. Considering the non-spherical nature of the particles, we proposed four additional parameters to describe the flow problem: two to parametrize the mutual orientation of the non-spherical particles ($S_1$ and $S_2$) and two to represent the polar and azimuthal angles ($\alpha$ and $\beta$) of the averaged flow velocity with respect to the configuration. For this, we have developed different static particle configurations using Monte-Carlo simulations. In the process, the configurations are biased to the desired amount of nematic or bi-axial orientational order with the use of a Maier-Saupe potential. The flow simulations indicate that the particle forces are configuration independent, at least for $\epsilon_s \le 0.4$, implying that the four additional parameters do not influence the results. The only important parameter representing orientation dependence is the incident angle $\phi$ of individual particles with respect to the average flow direction. 

The configuration independence greatly simplifies the parameter space to be explored from 6 to 3 dimensions, namely $\Rey$, $\epsilon_s$, and $\phi$. Of the three, the simulations are set up for only two parameters: $\Rey$ and $\epsilon_s$. Given a sufficiently random particle configuration, different incident angles $\phi$ are covered automatically. Another interesting result from the current work is that our previous finding of sine-squared scaling of drag for isolated non-spherical particles \citep{sanjeevi2017orientational} applies also to multiparticle systems. In other words, given a $\Rey$ and $\epsilon_s$, the drag on the subset of particles oriented at an incident angle $\phi$ with respect to the superficial flow velocity can be described with the knowledge of drag at $\phi=0^\circ$ and $\phi=90^\circ$ alone. This information can be used in a packed bed to determine the pressure drop across the bed with the knowledge of $\phi$ distribution alone. In a multiparticle configuration, also the lift on a particle at an incident angle $\phi$ can be computed with good accuracy using the drag at $\phi=0^\circ$ and $\phi=90^\circ$, as in our previous work on isolated non-spherical particles. Having identified the dependent parameters, we proposed correlations for drag, lift and torque for a multiparticle configuration. During the process, we used correlations for isolated non-spherical particles and extended them to the multiparticle systems.

We have also explored the validity of the conventional approach of combining known correlations for isolated non-spherical particle drag with correlations for voidage effects based on sphere packings. We observe that in the dilute and intermediate $\epsilon_s$ regimes ($\epsilon_s\leq0.3$), the influence of $\epsilon_s$ is nearly shape independent. This implies that the above conventional approach can safely be used to mimic flow around assemblies of non-spherical particles upto intermediate $\epsilon_s$. However for denser regimes, there is a need for multiparticle simulations and hence the need for this work. In the inertial regimes, the ratios of drag at $\phi=90^\circ$ and $\phi=0^\circ$ ($F_{D,\phi=90^\circ}/F_{D,\phi=0^\circ}$) are nearly constant until $\epsilon\leq0.3$ and then decrease with increasing $\epsilon_s$. This further proves that the conventional approach is not valid for dense regimes. In the process, we have analysed the flow velocity distribution as function of $\Rey$ and $\epsilon_s$. Likewise, the influence of different particle configurations on the flow velocities have also been analysed.

Overall, this work provides a recipe to parametrize the drag, lift and torque experienced by non-spherical particles in multiparticle environment. To the best of the authors' knowledge, there exists no work which parametrizes the drag, lift and torque for non-spherical particles in a multiparticle environment. Generally, lift and torque are ignored in large scale Euler-Lagrangian simulations. The proposed accurate drag, lift and torque correlations enable future Euler-Lagrangian simulations to be performed with more realistic physics.

\section*{Acknowledgements}
The authors thank the European Research Council for its financial support under its consolidator grant scheme, contract no. 615096 (NonSphereFlow). The simulations are performed in LB3D code with improved boundary conditions. We thank Prof. J. Harting for the code framework. SKPS thanks Y. El Hasadi for the fruitful discussions on fitting the data and I. Mema for the data in figure \ref{fig:packedBed}. The work has been made possible by a grant for computation time, project number SH-351-15, financed by the Netherlands Organisation for Scientific Research (NWO).




\bibliographystyle{jfm}
\bibliography{reference}

\begin{thebibliography}{28}
\expandafter\ifx\csname natexlab\endcsname\relax\def\natexlab#1{#1}\fi
\def\au#1{#1} \def\ed#1{#1} \def\yr#1{#1}\def\at#1{#1}\def\jt#1{\textit{#1}}
  \def\bt#1{#1}\def\bvol#1{\textbf{#1}} \def\vol#1{#1} \def\pg#1{#1}
  \def\publ#1{#1}\def\arxiv#1{#1}\def\org#1{#1}\def\st#1{\textit{#1}}

\bibitem[Beetstra {\em et~al.\/}(2007)Beetstra, van~der Hoef \&
  Kuipers]{beetstra2007drag}
{\sc \au{Beetstra, R.}, \au{van~der Hoef, M.~A.} \& \au{Kuipers, J. A.~M.}}
  \yr{2007}  \at{{Drag force of intermediate Reynolds number flow past mono-and
  bidisperse arrays of spheres}}.  \jt{AIChE J.}  \bvol{53}~(2),
  \pg{489--501}.

\bibitem[Bouzidi {\em et~al.\/}(2001)Bouzidi, Firdaouss \&
  Lallemand]{bouzidi2001momentum}
{\sc \au{Bouzidi, M.}, \au{Firdaouss, M.} \& \au{Lallemand, P.}} \yr{2001}
  \at{{Momentum transfer of a Boltzmann-lattice fluid with boundaries}}.
  \jt{Phys. Fluids}  \bvol{13},  \pg{3452--3459}.

\bibitem[d'Humi{\`e}res {\em et~al.\/}(2002)d'Humi{\`e}res, Ginzburg, Krafczyk,
  Lallemand \& Luo]{d2002multiple}
{\sc \au{d'Humi{\`e}res, D.}, \au{Ginzburg, I.}, \au{Krafczyk, M.},
  \au{Lallemand, P.} \& \au{Luo, L.-S.}} \yr{2002}
  \at{{Multiple--relaxation--time lattice Boltzmann models in three
  dimensions}}.  \jt{Phil. Trans. R. Soc. Lond. A}  \bvol{360},  \pg{437--451}.

\bibitem[Di~Felice(1994)]{di1994voidage}
{\sc \au{Di~Felice, R.}} \yr{1994}  \at{{The voidage function for
  fluid-particle interaction systems}}.  \jt{Int. J. Multiph. Flow}
  \bvol{20}~(1),  \pg{153--159}.

\bibitem[Ergun(1952)]{ergun1952fluid}
{\sc \au{Ergun, S.}} \yr{1952}  \at{{Fluid flow through packed columns}}.
  \jt{Chem. Eng. Prog.}  \bvol{48},  \pg{89--94}.

\bibitem[He \& Tafti(2018)]{he2018variation}
{\sc \au{He, L.} \& \au{Tafti, D.}} \yr{2018}  \at{{Variation of drag, lift and
  torque in a suspension of ellipsoidal particles}}.  \jt{Powder Technol.}
  \bvol{335},  \pg{409--426}.

\bibitem[Hill {\em et~al.\/}(2001)Hill, Koch \& Ladd]{hill2001moderate}
{\sc \au{Hill, R.~J.}, \au{Koch, D.~L.} \& \au{Ladd, A. J.~C.}} \yr{2001}
  \at{{Moderate-Reynolds-number flows in ordered and random arrays of
  spheres}}.  \jt{J. Fluid Mech.}  \bvol{448},  \pg{243--278}.

\bibitem[van~der Hoef {\em et~al.\/}(2005)van~der Hoef, Beetstra \&
  Kuipers]{van2005lattice}
{\sc \au{van~der Hoef, M.~A.}, \au{Beetstra, R.} \& \au{Kuipers, J. A.~M.}}
  \yr{2005}  \at{{Lattice-Boltzmann simulations of low-Reynolds-number flow
  past mono-and bidisperse arrays of spheres: results for the permeability and
  drag force}}.  \jt{J. Fluid Mech.}  \bvol{528},  \pg{233--254}.

\bibitem[Holloway {\em et~al.\/}(2010)Holloway, Yin \&
  Sundaresan]{holloway2010fluid}
{\sc \au{Holloway, W.}, \au{Yin, X.} \& \au{Sundaresan, S.}} \yr{2010}
  \at{{Fluid-particle drag in inertial polydisperse gas--solid suspensions}}.
  \jt{AIChE J.}  \bvol{56}~(8),  \pg{1995--2004}.

\bibitem[H{\"o}lzer \& Sommerfeld(2008)]{holzer2008new}
{\sc \au{H{\"o}lzer, A.} \& \au{Sommerfeld, M.}} \yr{2008}  \at{{New simple
  correlation formula for the drag coefficient of non-spherical particles}}.
  \jt{Powder Technol.}  \bvol{184}~(3),  \pg{361--365}.

\bibitem[Huang {\em et~al.\/}(2012)Huang, Yang, Krafczyk \&
  Lu]{huang2012rotation}
{\sc \au{Huang, H.}, \au{Yang, X.}, \au{Krafczyk, M.} \& \au{Lu, X.-Y.}}
  \yr{2012}  \at{{Rotation of spheroidal particles in Couette flows}}.  \jt{J.
  Fluid Mech.}  \bvol{692},  \pg{369--394}.

\bibitem[Lallemand \& Luo(2003)]{lallemand2003lattice}
{\sc \au{Lallemand, P.} \& \au{Luo, L.-S.}} \yr{2003}  \at{{Lattice Boltzmann
  method for moving boundaries}}.  \jt{J. Comput. Phys.}  \bvol{184},
  \pg{406--421}.

\bibitem[Mahajan {\em et~al.\/}(2018)Mahajan, Padding, Nijssen, Buist \&
  Kuipers]{mahajan2018nonspherical}
{\sc \au{Mahajan, V.~V.}, \au{Padding, J.~T.}, \au{Nijssen, T. M.~J.},
  \au{Buist, K.~A.} \& \au{Kuipers, J. A.~M.}} \yr{2018}  \at{{Nonspherical
  particles in a pseudo-2D fluidized bed: Experimental study}}.  \jt{AIChE J.}
  \bvol{64}~(5),  \pg{1573--1590}.

\bibitem[Maier \& Saupe(1959)]{maier1959einfache}
{\sc \au{Maier, W.} \& \au{Saupe, A.}} \yr{1959}  \at{{Eine einfache
  molekular-statistische Theorie der nematischen kristallinfl{\"u}ssigen Phase.
  Teil 1.}}  \jt{Z. Naturforsch. A}  \bvol{14}~(10),  \pg{882--889}.

\bibitem[Oschmann {\em et~al.\/}(2014)Oschmann, Hold \&
  Kruggel-Emden]{oschmann2014numerical}
{\sc \au{Oschmann, T.}, \au{Hold, J.} \& \au{Kruggel-Emden, H.}} \yr{2014}
  \at{{Numerical investigation of mixing and orientation of non-spherical
  particles in a model type fluidized bed}}.  \jt{Powder Technol.}  \bvol{258},
   \pg{304--323}.

\bibitem[Pournin {\em et~al.\/}(2005)Pournin, Weber, Tsukahara, Ferrez,
  Ramaioli \& Liebling]{pournin2005three}
{\sc \au{Pournin, L.}, \au{Weber, M.}, \au{Tsukahara, M.}, \au{Ferrez, J.-A.},
  \au{Ramaioli, M.} \& \au{Liebling, T.~M.}} \yr{2005}  \at{{Three-dimensional
  distinct element simulation of spherocylinder crystallization}}.  \jt{Granul.
  Matter}  \bvol{7}~(2-3),  \pg{119--126}.

\bibitem[Richardson \& Zaki(1954)]{richardson1954sedimentation}
{\sc \au{Richardson, J.~F.} \& \au{Zaki, W.~N.}} \yr{1954}  \at{{Sedimentation
  and fluidization: Part 1}}.  \jt{Trans. Inst. Chem. Eng}  \bvol{32},
  \pg{35--53}.

\bibitem[Richter \& Nikrityuk(2013)]{richter2013new}
{\sc \au{Richter, A.} \& \au{Nikrityuk, P.~A.}} \yr{2013}  \at{{New
  correlations for heat and fluid flow past ellipsoidal and cubic particles at
  different angles of attack}}.  \jt{Powder Technol.}  \bvol{249},
  \pg{463--474}.

\bibitem[Rubinstein {\em et~al.\/}(2017)Rubinstein, Ozel, Yin, Derksen \&
  Sundaresan]{rubinstein2017lattice}
{\sc \au{Rubinstein, G.~J.}, \au{Ozel, A.}, \au{Yin, X.}, \au{Derksen, J.~J.}
  \& \au{Sundaresan, S.}} \yr{2017}  \at{{Lattice Boltzmann simulations of
  low-Reynolds-number flows past fluidized spheres: effect of inhomogeneities
  on the drag force}}.  \jt{J. Fluid Mech.}  \bvol{833},  \pg{599--630}.

\bibitem[Sanjeevi {\em et~al.\/}(2018{\natexlab{{\em a\/}}})Sanjeevi, Kuipers
  \& Padding]{sanjeevi2018drag}
{\sc \au{Sanjeevi, S. K.~P.}, \au{Kuipers, J. A.~M.} \& \au{Padding, J.~T.}}
  \yr{2018{\natexlab{{\em a\/}}}}  \at{{Drag, lift and torque correlations for
  non-spherical particles from Stokes limit to high Reynolds numbers}}.
  \jt{Int. J. Multiph. Flow}  \bvol{106},  \pg{325--337}.

\bibitem[Sanjeevi \& Padding(2017)]{sanjeevi2017orientational}
{\sc \au{Sanjeevi, S. K.~P.} \& \au{Padding, J.~T.}} \yr{2017}  \at{{On the
  orientational dependence of drag experienced by spheroids}}.  \jt{J. Fluid
  Mech.}  \bvol{820}.

\bibitem[Sanjeevi {\em et~al.\/}(2018{\natexlab{{\em b\/}}})Sanjeevi, Zarghami
  \& Padding]{sanjeevi2018choice}
{\sc \au{Sanjeevi, S. K.~P.}, \au{Zarghami, A.} \& \au{Padding, J.~T.}}
  \yr{2018{\natexlab{{\em b\/}}}}  \at{{Choice of no-slip curved boundary
  condition for lattice Boltzmann simulations of high-Reynolds-number flows}}.
  \jt{Phys. Rev. E}  \bvol{97}~(4),  \pg{043305}.

\bibitem[Schiller \& Naumann(1935)]{Schiller1935drag}
{\sc \au{Schiller, L.} \& \au{Naumann, A.}} \yr{1935}  \at{{A drag coefficient
  correlation}}.  \jt{Z. Ver. Deutsch. Ing}  \bvol{77},  \pg{318--320}.

\bibitem[Tang {\em et~al.\/}(2015)Tang, Peters, Kuipers, Kriebitzsch \& van~der
  Hoef]{yali2015new}
{\sc \au{Tang, Y.}, \au{Peters, E. A. J.~F.}, \au{Kuipers, J. A.~M.},
  \au{Kriebitzsch, S. H.~L.} \& \au{van~der Hoef, M.~A.}} \yr{2015}  \at{{A new
  drag correlation from fully resolved simulations of flow past monodisperse
  static arrays of spheres}}.  \jt{AIChE J.}  \bvol{61}~(2),  \pg{688--698}.

\bibitem[Tenneti {\em et~al.\/}(2011)Tenneti, Garg \&
  Subramaniam]{tenneti2011drag}
{\sc \au{Tenneti, S.}, \au{Garg, R.} \& \au{Subramaniam, S.}} \yr{2011}
  \at{{Drag law for monodisperse gas--solid systems using particle-resolved
  direct numerical simulation of flow past fixed assemblies of spheres}}.
  \jt{Int. J. Multiph. Flow}  \bvol{37}~(9),  \pg{1072--1092}.

\bibitem[Vega \& Lago(1994)]{vega1994fast}
{\sc \au{Vega, C.} \& \au{Lago, S.}} \yr{1994}  \at{{A fast algorithm to
  evaluate the shortest distance between rods}}.  \jt{Comput. Chem. (Oxford)}
  \bvol{18}~(1),  \pg{55--59}.

\bibitem[Williams \& Philipse(2003)]{williams2003random}
{\sc \au{Williams, S.~R.} \& \au{Philipse, A.~P.}} \yr{2003}  \at{{Random
  packings of spheres and spherocylinders simulated by mechanical
  contraction}}.  \jt{Phys. Rev. E}  \bvol{67}~(5),  \pg{051301}.

\bibitem[Zastawny {\em et~al.\/}(2012)Zastawny, Mallouppas, Zhao \&
  Van~Wachem]{zastawny2012derivation}
{\sc \au{Zastawny, M.}, \au{Mallouppas, G.}, \au{Zhao, F.} \& \au{Van~Wachem,
  B.}} \yr{2012}  \at{{Derivation of drag and lift force and torque
  coefficients for non-spherical particles in flows}}.  \jt{Int. J. Multiph.
  Flow}  \bvol{39},  \pg{227--239}.

\end{thebibliography}
\end{document}